\documentclass[prc,a4paper,showpacs,byrevtex]{revtex4}
\usepackage{graphicx}
\usepackage{dcolumn}
\usepackage{amsmath}
\usepackage{array}
\usepackage{bm}
\usepackage{amssymb}
\usepackage{amsfonts}

\begin{document}

\title[${\cal N}$-body problem and auxiliary field method]
{The quantum ${\cal N}$-body problem and the auxiliary field method}

\author{Bernard Silvestre-Brac}
\email[E-mail: ]{silvestre@lpsc.in2p3.fr}
\affiliation{LPSC Universit\'{e} Joseph Fourier, Grenoble 1,
CNRS/IN2P3, Institut Polytechnique de Grenoble, 
Avenue des Martyrs 53, F-38026 Grenoble-Cedex, France}
\author{Claude Semay}
\email[E-mail: ]{claude.semay@umh.ac.be}
\author{Fabien Buisseret}
\email[E-mail: ]{fabien.buisseret@umh.ac.be}
\author{Fabian Brau}
\email[E-mail: ]{fabian.brau@umh.ac.be}
\affiliation{Service de Physique Nucl\'{e}aire et Subnucl\'{e}aire, Universit\'{e}
de Mons - UMONS, 20 Place du Parc,
7000 Mons, Belgium}

\date{\today}

\begin{abstract}
Approximate analytical energy formulas for ${\cal N}$-body semirelativistic Hamiltonians with one- and two-body interactions are obtained within the framework of the auxiliary field method. This method has already been proved to be a powerful technique in the case of two-body problems. A general procedure is given and applied to various Hamiltonians of interest, in atomic and hadronic physics in particular. A test of formulas is performed for baryons described as a three-quark system.
\end{abstract}

\maketitle

\section{Introduction}

The quantum-mechanical ${\cal N}$-body problem has deserved a tremendous amount of work since more than 50 years. Both numerical and analytical resolutions of that problem are highly nontrivial as soon as ${\cal N}>2$. Actually, even restricting to two-body systems, an analytical solution is generally not known although numerical computations are quite straightforward in this case. Because quantum ${\cal N}$-body problems arise in all areas where quantum mechanics is involved, from atomic to hadronic  physics, the amount of papers devoted to that topic is huge. We refer the reader to several textbooks in which many analytical results as well as useful references can be found~\cite{flu,nbor,kinetic}.  

Recently, we have proposed a new method, called auxiliary field method (AFM), to compute the bound states of a given two-body Hamiltonian \cite{af,af2,af3,afmenv,hybri,afrela}. This method is based on auxiliary (einbein) fields and can lead to analytical approximate closed formulas for various central interactions, with both nonrelativistic or relativistic kinematics. The dependence of mass formulas on parameters of the Hamiltonian can generally be determined from a simple dimensional analysis. But one of the interesting features of the AFM is to predict, at least approximately, the dependence on quantum numbers.

The purpose of the present paper is to show that the AFM can be successfully applied to find approximate analytical mass formulas for general ${\cal N}$-body Hamiltonians of the form
\begin{equation}\label{hamgene}
	H=\sum^{{\cal N}}_{i=1}\sqrt{\bm p^2_i+m^2_i}+\sum^{{\cal N}}_{i=1} V_i(|\bm r_i-\bm R|)+\sum^{{\cal N}}_{i<j=1}\bar V_{ij}(|\bm r_i-\bm r_j|),
\end{equation}
with $\sum^{{\cal N}}_{i=1} \bm p_i=\bm P=\bm 0$, where $\bm r_i$ and $\bm p_i$ are respectively the position and momentum of particle $i$ with a mass $m_i$, and where $\bm R$ is a global variable to be specified later. In the following, we will use the natural units $\hbar=c=1$. We consider a relativistic kinetic energy and allow for both pairwise and one-body interactions. Such a Hamiltonian being not Lorentz invariant, it is often called ``semirelativistic" in the literature, see \textit{e.g.}~\cite{hallsr1,hallsr2}. Particular systems exhibiting one-body potentials of the form appearing in Eq.~(\ref{hamgene}) will be studied below.  

Our paper is organized as follows. We begin by showing how the ground-state energy of a ${\cal N}$-body Hamiltonian with two-body interactions can be estimated from the corresponding two-body Hamiltonian in Sec.~\ref{twobr}. Then, we focus on the ${\cal N}$-body harmonic oscillator in Sec.~\ref{nboh}. This problem is particularly interesting since it can be fully analytically solved without resorting to the AFM in the nonrelativistic case, while the AFM is needed in the semirelativistic case. The harmonic oscillator is the starting point of the general application of the AFM to the generic Hamiltonian~(\ref{hamgene}). The procedure is discussed in Sec.~\ref{afm}, followed by several examples: power-law potentials in Sec.~\ref{sec:plpot}, baryonic and atomic systems in Secs.~\ref{baryon} and \ref{atom}. The case for which only two-body interactions are present is discussed in Sec.~\ref{pair}. Finally, an application of our formulas is given in Sec.~\ref{appbar} and conclusions are drawn in Sec.~\ref{conclu}.

\section{Two-body reduction}\label{twobr}

Finding exact results for two-body Hamiltonians is obviously a less problematic task than in the ${\cal N}$-body case, not only analytically but also numerically. That is why it is of interest to try to reduce ${\cal N}$-body Hamiltonians to equivalent two-body problems. That idea is already present in early works about nuclear physics such as Ref.~\cite{wigner}. Further references about that technique can be found in Ref.~\cite{hall2b}. In the same spirit, we show in this section how the AFM can be applied to express the ground-state energy of a ${\cal N}$-body system in terms of the one of a two-body system.  

\subsection{General results}

Several analytical approximate energy formulas have already been found by applying the AFM to one- or two-body Hamiltonians, see Refs.~\cite{af,af2,af3,afmenv,hybri,afrela}. These formulas can actually be used to compute the \textbf{ground-state mass} of a system of ${\cal N}$
\textbf{identical} particles interacting \textit{via} pairwise interactions. Let us start from the ${\cal N}$-body spinless Salpeter Hamiltonian    
\begin{equation}\label{nbod0}
	H^{({\cal N})}=\sum^{{\cal N}}_{i=1}\sqrt{\bm p^2_i+m^2}+\sum^{\cal N}_{i<j=1}\bar V(r_{ij}), \quad {\rm where}\quad r_{ij}=|\bm r_i-\bm r_j|,
\end{equation}
which is a particular case of Hamiltonian~(\ref{hamgene}). We denote $M^{({\cal N})}(m)$ the exact ground state mass of the system governed by the
Hamiltonian (\ref{nbod0}); we want an approximate analytical expression of this quantity.

Finding analytical solutions for a wide range of potentials would be hopeless because of the square roots of the form $\sqrt{\bm p^2+m^2}$ appearing in the kinetic term. Nevertheless, as already pointed out in Ref.~\cite{martinsr}, the operator inequality
\begin{equation}\label{ineq}
\sqrt{\bm p^2+m^2}<\frac{\bm p^2+m^2+\mu^2}{2\mu},
\end{equation}
where $\mu$ is a real number, can be used to obtain upper bounds on the exact semirelativistic Hamiltonian. Analytical results can then be hoped since the introduction of the variational parameter $\mu$ leads to an apparently nonrelativistic kinetic energy. Such a technique has been further developed in Refs.~\cite{hallsr1,Luchasr,hallsr3}, where explicit example are also given. In our framework, the auxiliary fields, denoted as $\nu_j$, play the role of the parameter $\mu$. When introduced in Hamiltonian~(\ref{nbod0}) in order to get rid of the square roots $\sqrt{\bm p^2_i+m^2}$, one obtains 
\begin{equation}\label{nbod1}
	H^{({\cal N})}(\nu_j)=\sum^{{\cal N}}_{i=1}\left[\frac{{\cal N}}{4}\frac{\bm p^2_i+m^2}{\nu_i}+\frac{\nu_i}{{\cal N}}\right]+\sum^{\cal N}_{i<j=1}\bar V(r_{ij}).
\end{equation}
We refer the reader to Ref.~\cite{afrela} for a detailed discussion about this procedure, but recall here the main ideas for completeness. The auxiliary fields are \textit{a priori} operators that can be eliminated by solving the constraints $\left.\delta_{\nu_i}H^{({\cal N})}(\nu_j)\right|_{\nu_i=\hat\nu_i}=0$, which are the Euler-Lagrange equation for the field $\nu_i$. It can then be checked that $\hat\nu_i={\cal N}\sqrt{\bm p^2_i+m^2}/2$ and that $H^{({\cal N})}(\hat\nu_j)=H^{({\cal N})}$. Both Hamiltonians~(\ref{nbod0}) and (\ref{nbod1}) are equivalent up to a proper elimination of the auxiliary fields. The computations are however considerably simplified if one makes the approximation that the auxiliary fields are real numbers. Then, Hamiltonian~(\ref{nbod1}) becomes formally a nonrelativistic one, and the auxiliary fields have to be seen as variational parameters, whose optimal values, $\nu_{j,0}$, are eventually obtained by minimizing the eigenvalues of~(\ref{nbod1}), denoted as $M^{({\cal N})}(\nu_j;m)$. Provided that Hamiltonian~(\ref{nbod1}) is analytically solvable or that an upper bound of its eigenvalues can be found, the inequality~(\ref{ineq}) says that $M^{({\cal N})}(\nu_{j,0};m)$ is an upper bound of the exact mass $M^{({\cal N})}(m)$, that is
\begin{equation}\label{inegmas1}
M^{({\cal N})}(m) \leq M^{({\cal N})}(\nu_{j,0};m) \leq M^{({\cal N})}(\nu_j;m).
\end{equation}
The optimal values $\nu_{j,0}$ depend on quantum numbers of the state considered. If a lower bound of the eigenvalues of~(\ref{nbod1}) is found however, no conclusion can be drawn about the relative position of this last bound and $M^{({\cal N})}(m)$.

From the AFM, we know that the optimal values $\nu_{j,0}$ should be close to $\left\langle \hat\nu_j\right\rangle$~\cite{af,afrela}. The ground state being the most symmetrical state, it is reasonable to assume that $\nu_j=\nu$ by symmetry, \textit{i.e.} that the total kinetic energy is equally distributed in average between the different particles. This will shown in Sec.~\ref{idpart}. With such a choice, the AFM can be applied with a single auxiliary field $\nu$ instead of ${\cal N}$ auxiliary fields $\nu_j$, a much simpler situation. Consequently one is led to 
\begin{equation}\label{nbod1b}
H^{({\cal N})}(\nu)=\frac{{\cal N}^2m^2}{4\nu}+\nu+\frac{{\cal N}}{4\nu}\sum^{{\cal N}}_{i=1}\bm p^2_i+\sum^{\cal N}_{i<j=1}\bar V(r_{ij}).
\end{equation}
Let us denote by $M^{({\cal N})}(\nu;m)$ the ground state mass of the simpler Hamiltonian (\ref{nbod1b}) and $\nu_0$ the value that minimizes this quantity. Due to the fact that the mass depends now on a single auxiliary field, the expression of $M^{({\cal N})}(\nu_0;m)$ is much easily obtained than $M^{({\cal N})}(\nu_{j,0};m)$. As $M^{({\cal N})}(\nu;m)=M^{({\cal N})}(\nu_j=\nu;m)$, we can write
\begin{equation}\label{inegmas2}
M^{({\cal N})}(m) \leq M^{({\cal N})}(\nu_0;m) \leq M^{({\cal N})}(\nu;m).
\end{equation}

Let us consider the two-body nonrelativistic problem of two particles of mass $\nu$ interacting through the potential $\bar V(r)$. The corresponding Hamiltonian is thus $\bm p^2/\nu +\bar V(r)$ whose ground state energy is $\epsilon(\nu)$.
It is convenient to use the following identity~\cite{kinetic}
\begin{equation}
\label{chgvar}
{\cal N} \sum^{{\cal N}}_{i=1}\bm p^2_i=4\sum^{\cal N}_{i<j=1}\bm \pi^2_{ij}+\left(\sum^{{\cal N}}_{i=1}\bm p_i\right)^2,
\end{equation}
where $\bm \pi_{ij}=\frac{\bm p_i-\bm p_j}{2}$ is the relative momentum of the pair $(ij)$. Since we are working in a frame where $\sum^{{\cal N}}_{i=1}\bm p_i=\bm 0$, the second term in the right-hand side of (\ref{chgvar}) cancels and Hamiltonian~(\ref{nbod1}) becomes exactly
\begin{equation} \label{hamsep}
		H^{({\cal N})}(\nu)=\frac{{\cal N}^2m^2}{4\nu}+\nu+\sum^{\cal N}_{i<j=1}\left[\frac{\bm \pi^2_{ij}}{\nu}+\bar V(r_{ij})\right].
\end{equation}
Let us assume that we know an analytical expression of $\epsilon(\nu)$. Since we are dealing with a system of ${\cal N}$ particles in their ground state, and since the Hamiltonian (\ref{hamsep}) is separable, it can be shown that a lower bound on the ground-state mass $M^{({\cal N})}(\nu)$ of Hamiltonian (\ref{hamsep}) is given by~\cite{kinetic} 
\begin{equation}\label{nbod3}
M^{({\cal N})}(\nu;m)\geq M^{({\cal N})}_0(\nu;m)= \frac{{\cal N}^2m^2}{4\nu}+\nu+\frac{{\cal N}({\cal N}-1)}{2} \epsilon(\nu).
\end{equation}

Now, let us consider the semirelativistic version of the two-body problem, whose Hamiltonian is given by $H^{(2)}=\sigma\, \sqrt{\bm p^2+m^2}+\bar V(r)$. In principle the two-body case implies $\sigma = 2$. However, for further convenience, $\sigma$ is chosen as an arbitrary positive real number. The exact ground state mass of this Hamiltonian is denoted $M^{(2)}(m,\sigma)$.

Introducing the auxiliary field $\mu$, one can use AFM to have an approximate expression of this quantity. Explicitly the auxiliary Hamiltonian is
\begin{equation}
\label{Hgnu}
	H^{(2)}(\mu)=\frac{m^2}{\mu}+\frac{\sigma^2\mu}{4}+\frac{\bm p^2}{\mu}+\bar V(r).
\end{equation}
The corresponding ground state $M^{(2)}(\mu;m,\sigma)$ is given by
\begin{equation}
\label{M2mu}
 M^{(2)}(\mu;m,\sigma)=\frac{m^2}{\mu}+\frac{\sigma^2\mu}{4}+\epsilon(\mu).
\end{equation}
One has the following inequalities
\begin{equation}
\label{ineqM2}
M^{(2)}(m,\sigma) \leq M^{(2)}(\mu_0;m,\sigma)  \leq M^{(2)}(\mu;m,\sigma),
\end{equation}
where $\mu_0$ is the value minimizing $M^{(2)}(\mu;m,\sigma)$.
Comparing (\ref{nbod3}) to (\ref{M2mu}), one can easily show that
\begin{equation}
M^{({\cal N})}_0(\nu;m)=\frac{{\cal N}({\cal N}-1)}{2} M^{(2)}\left(\nu;m'=m\sqrt{\frac{{\cal N}}{2({\cal N}-1)}},\sigma'=\sqrt{\frac{8}{{\cal N}({\cal N}-1)}}\right).
\end{equation}
Consequently, we arrive at the following series of inequalities $M^{({\cal N})}(\nu_0;m) \geq M^{({\cal N})}_0(\nu_0;m) = ({\cal N}({\cal N}-1)/2)
M^{(2)}(\nu_0;m',\sigma') \geq ({\cal N}({\cal N}-1)/2)M^{(2)}(\mu_0;m',\sigma') \geq ({\cal N}({\cal N}-1)/2)M^{(2)}(m',\sigma')$.

The previous inequalities and inequalities (\ref{inegmas2}) are summarized by
\begin{eqnarray}
\label{dlineq}
M^{({\cal N})}(m) &\leq & M^{({\cal N})}(\nu_0;m), \\
\frac{{\cal N}({\cal N}-1)}{2}M^{(2)}(m',\sigma') &\leq& M^{({\cal N})}(\nu_0;m).
\end{eqnarray}
If $M^{({\cal N})}(\nu_0;m)$ represents a good approximation of the ${\cal N}$-body energy, it seems reasonable to assume that the values of the left hand parts should be rather close. We arrive at the final approximate formula for the ${\cal N}$-body ground-state mass in terms of the two-body ground-sate mass
\begin{equation}\label{nbod5}
M^{({\cal N})}(m)  \approx \frac{{\cal N}({\cal N}-1)}{2}\ M^{(2)}\left(m\sqrt{\frac{{\cal N}}{2({\cal N}-1)}},\sqrt{\frac{8}{{\cal N}({\cal N}-1)}}\right).
\end{equation}
This means that all the ground state mass formulas that can be obtained in the two-body case can be generalized to the ground state of a ${\cal N}$-particle system thanks to (\ref{nbod5}). It is worth pointing out that the result~(\ref{nbod5}) has been firstly conjectured in Ref.~\cite{hall1} and explicitly proved for ${\cal N}=3$, 4 in this last reference. It is an interesting check of the AFM that we recover this previously established result. For completeness, we mention that other lower bounds can be found for example in Refs.~\cite{hall2,hall3}.   

Let us emphasize an important point. In this section the auxiliary field $\nu$ was introduced to deal only with the worrying square root appearing in the kinetic energy operator. This means that, in order to apply (\ref{nbod5}), one must be able to obtain the exact eigenvalues of the Hamiltonian $H^{(2)}(\nu)$ [see Eq.~(\ref{Hgnu})] with the given potential $\bar V(r)$. In that case, the spectrum obtained is an upper bound of the exact one. Indeed, very rare are the situations for which this procedure can be achieved. More frequently, one has only approximate expressions for $\epsilon(\nu)$; in this respect the AFM can be invoked once more, but at the price of introducing supplementary auxiliary fields \cite{af}. Then, an upper bound of the spectrum is obtained if an upper bound of $\epsilon(\nu)$ can be found, and no conclusion can be drawn if a lower bound of $\epsilon(\nu)$ is obtained. This approach, dealing with the kinetic energy and potential operators on equal footing, will be discussed in further sections.

\subsection{An example}

As an example of the above considerations, let us consider the case of ${\cal N}$ identical particles interacting \textit{via} a Coulomb potential $-b/r_{ij}$. Applying the AFM, the ground-state mass of Hamiltonian $\sigma\sqrt{\bm p^2+m^2}-b/r$ is given by~\cite{afrela}
\begin{equation}
M_{g;{\rm c}}(m,\sigma)	=\sigma m \sqrt{1-\frac{b^2}{\sigma^2Q^2_{{\rm c}}|_{n=l=0}}}.
\end{equation}
In this last relation, $Q_{\rm c}=n+l+1$ \textit{a priori}, but it can also be replaced by a more complicated function of the form $Q_{\rm c}=\alpha(b)\, n+l+\beta(b)$, with $\alpha$ and $\beta$ continuous functions of $b$, giving better results~\cite{afrela}. Equation~(\ref{nbod5}) leads then to
\begin{equation}\label{mcoul}
M^{({\cal N})}_{g;{\rm c}}\approx {\cal N}m\sqrt{1-\frac{{\cal N}({\cal N}-1)}{8}\frac{b^2}{Q^2_{\rm c}|_{n=l=0}}} .
\end{equation}
With a relativistic kinematics, the strength of the Coulomb potential must not be too strong otherwise the Hamiltonian is not bounded from below and the system collapses. It is easy to notice that the critical value, $b_{{\cal N}}$, for the ${\cal N}$-body problem is related to the critical value, $b_2$, for the two-body problem through
\begin{equation}\label{acritn2}
b_{{\cal N}}\approx \sqrt{\frac{2}{{\cal N}({\cal N}-1)}} \; b_2.
\end{equation}
Thus it decreases as the number of interacting pairs increases. We also recover a result given in Ref.~\cite{hall4,lieb87}: At large ${\cal N}$, the stability of the system is guaranteed if $b=\tilde b/{\cal N}$, where $\tilde b$ is a constant below some critical value.

\section{The ${\cal N}$-body harmonic oscillator}\label{nboh}

The case of a quadratic potential, \textit{i.e.} the ${\cal N}$-body harmonic oscillator, is practically the only one for which an exact solution is reachable. For that reason, it is the basic ingredient of our method when dealing with ${\cal N}$-body Hamiltonians. The ${\cal N}$-body harmonic oscillator problem has been widely studied in the literature, especially in the nonrelativistic case. A first analytical solution of that problem can be found in Ref.~\cite{houston} together with an application to nuclear physics, while the exact spectrum of a nonrelativistic ${\cal N}$-body harmonic oscillator with pairwise interactions has been given in Ref.~\cite{hallho}. The interested reader will find a detailed list of references concerning that topic in Sec. III of Ref.~\cite{hallho2}, as well as a discussion of the changes caused by the consideration of fermions rather than bosons. In the semirelativistic case, upper and lower bounds are found in Ref.~\cite{hallsr2} for a ${\cal N}$-body harmonic oscillator with pairwise interactions.    

Before making explicit computations, it is worth mentioning the new results reported in this Section. First, we extend previous results in the nonrelativistic case by showing that exact eigenenergies can be found when both one- and two-body harmonic interactions are present [see Hamiltonian~(\ref{honr})]. Second, we give an explicit mass formula in the semirelativistc case with one- and two-body interactions, which is an upper bound of the exact mass formula. To our knowledge, that formula has never been obtained before.  

\subsection{Jacobi coordinates}

Usually the many-body problem is treated starting with the shell-model or variants. In this method, the degrees of freedom are simply the positions $\bm r_i$ of the various particles ($i = 1,\ldots,{\cal N}$). But, in this case, the motion of the center of mass is not separated correctly and this leads to spurious components which spoil the results. On the contrary, in the few-body problem, one introduces new degrees of freedom which allow to correct exactly this important drawback. The price to pay is that the resulting method becomes more and more difficult to be applied when the number of particles increases. In practice, one is limited to values less than 10. Fortunately, this method can be used in the case of the harmonic oscillator, whatever the value of ${\cal N}$, as we will see below.

Among the various possible new degrees of freedom, the Jacobi coordinates are of common use.  First, we choose a reference mass $m$ (it can be that of a particle or the total mass of the system for example) and define the dimensionless quantities 
\begin{equation}
\alpha_i=m_i/m, \quad \alpha_{12\dots i}=\sum^i_{k=1}\alpha_k\quad \textrm{and} \quad \alpha = \alpha_{12\ldots {\cal N}}. 
\end{equation}
Then, the standard Jacobi coordinates, $\bm x_i$, can be expressed as 
\begin{equation}
	\bm x_i=\frac{\sum^i_{k=1}\alpha_k\, \bm r_k}{\alpha_{12\dots i}}-\bm r_{i+1}, \quad i=1,\ldots,{\cal N}-1, \quad {\rm and}\quad \bm x_{{\cal N}}=\bm R = \frac{1}{\alpha} \sum^i_{k={\cal N}}\alpha_k\, \bm r_k.
\end{equation}
For convenience, the global coordinate $\bm R$ is relegated as the last Jacobi coordinate $\bm x_{{\cal N}}$ whereas $\bm x_i$ ($i=1$, \ldots, ${\cal N}-1$) represents the vector joining the nonrelativistic center of mass of the first $i$ particles to the particle $i+1$. In the case of nonrelativistic kinematics or in the case of a system composed of identical particles, $\bm R$ is the center of mass coordinate. By using the matrix notation $\bm x_i=\sum^{{\cal N}}_{j=1}U_{ij}\bm r_j$, one is led to the following definition of the $U$-matrix
\begin{subequations}
\begin{eqnarray}
U_{ij}&=&\frac{\alpha_j}{\alpha_{12\ldots i}} \quad {\rm if}\quad  j \leq i \\
U_{ii+1}&=&-1 \\
U_{ij}&=&0 \quad {\rm if}\quad  j > i+1.
\end{eqnarray}
\end{subequations}
It is not difficult to calculate the inverse matrix $B=U^{-1}$ whose matrix elements are given by
\begin{subequations}
\label{defB}
\begin{eqnarray}
B_{kl}&=&\frac{\alpha_{l+1}}{\alpha_{12\ldots l+1}} \quad {\rm if}\quad  k \leq l <  {\cal N} \\
B_{l+1 l}&=&- \frac{\alpha_{12\ldots l}}{\alpha_{12\dots l+1}} \quad {\rm if} \quad  l <  {\cal N} \\
B_{kl}&=&0 \quad {\rm if}\quad  k > l+1 \\
B_{k {\cal N}}&=&1 \quad \forall k.
\end{eqnarray}
\end{subequations}
Denoting $\bm p_i$ and $\bm \pi_i$ the conjugate momenta associated respectively to $\bm r_i$ and $\bm x_i$, it is easy to prove that
\begin{equation}
\bm \pi_i=\sum^{{\cal N}}_{j=1}B_{ji}\bm p_j \quad {\rm and} \quad \bm p_i=\sum^{{\cal N}}_{j=1}U_{ji}\bm \pi_j. 
\end{equation}
$\bm \pi_{\cal N} = \bm P$ is the total momentum of the system. In any case, we work in a frame for which $\bm \pi_{\cal N}=\bm 0$.  

\subsection{Nonrelativistic general case}

Let us start with the most general harmonic-oscillator-like Hamiltonian, corresponding to ${\cal N}$ particles of arbitrary masses, with one-body and two-body quadratic potentials with arbitrary positive string constants. The Hamiltonian looks like
\begin{equation}\label{honr}
	H_{{\rm ho}}=\sum^{{\cal N}}_{i=1}\frac{\bm p^2_i}{2m_i}+\sum^{{\cal N}}_{i=1} k_i (\bm r_i-\bm R)^2+\sum^{{\cal N}}_{i<j=1} \bar k_{ij} (\bm r_i-\bm r_j)^2.
\end{equation}
The kinetic energy operator, expressed in terms of Jacobi variables, allows the correct separation of the center of mass motion and appears decoupled in the various variables (this is in fact the justification of the form of Jacobi coordinates)
\begin{equation}
\label{Tcinoh}
T=\sum^{{\cal N}}_{i=1}\frac{\bm p^2_i}{2m_i} = \frac{\bm P^2}{2 m_t}+\sum^{{\cal N}-1}_{i=1} \frac{\lambda_i^2}{2m} \bm \pi_i^2,
\end{equation}
where $m_t = \alpha m=m_1+m_2+\cdots m_{\cal N}$ is the total mass of the system. The kinematical quantity $\lambda_i$ is calculated as
\begin{equation}
\label{deflambi}
\lambda_i = \left(\frac{\alpha_{12\ldots i+1}}{\alpha_{i+1} \alpha_{12\ldots i}}\right)^{1/2}.
\end{equation}
For further convenience, it is judicious to switch from the standard Jacobi variables to conjugate renormalized Jacobi coordinates defined as ($i=1$, \ldots, ${\cal N}-1$)
\begin{equation}
\bm y_i = \frac{\bm x_i}{\lambda_i}, \quad \bm \rho_i = \lambda_i \bm \pi_i.
\end{equation}
Working in the center of mass frame ($\bm P= \bm 0$) and using these new variables, the kinetic energy operator has a very simple form
\begin{equation}
T = \frac{1}{2m} \sum^{{\cal N}-1}_{i=1} \bm \rho_i^2.
\end{equation}
With the condition $\bm P= \bm 0$, it is clear that the case ${\cal N}=1$ cannot be treated by the following formulas. Results for this case are given in Ref.~\cite{afrela}.

With these new variables the one-body operator is written
\begin{equation} 
V_1 = \sum^{{\cal N}}_{i=1} k_i (\bm r_i-\bm R)^2 = \sum^{{\cal N}-1}_{l,m=1} F_{lm} \bm y_l \cdot \bm y_m,
\end{equation}
where the symmetric positive-definite matrix $F$ is defined by
\begin{equation}
\label{defF} 
F_{lm} = \lambda_l \lambda_m \sum^{{\cal N}}_{i=1} k_i B_{il} B_{im},
\end{equation}
the matrix $B$ being given by (\ref{defB}). In the very same way, the two-body operator is written
\begin{equation} 
V_2 = \sum^{{\cal N}}_{i<j=1} \bar k_{ij} (\bm r_i-\bm r_j)^2 = \sum^{{\cal N}-1}_{l,m=1} G_{lm} \bm y_l \cdot \bm y_m ,
\end{equation}
where the symmetric positive-definite matrix $G$ is defined by
\begin{equation}
\label{defG} 
G_{lm} = \lambda_l \lambda_m \sum^{{\cal N}}_{i<j=1} \bar k_{ij} (B_{il} - B_{jl}) (B_{im} - B_{jm}).
\end{equation}

Introducing the matrix $J=F+G$, the total potential $V=V_1+V_2$ is expressed as
\begin{equation} 
V = \sum^{{\cal N}-1}_{l,m=1} J_{lm} \bm y_l \cdot \bm y_m.
\end{equation}
For arbitrary masses or/and string constants there is no reason why the matrix $J$ should be diagonal. However, this matrix being a symmetric positive-definite matrix, it can be diagonalized with help of a unitary matrix (in fact an orthogonal one since all the quantities are real). Thus ($\tilde{O}$ denotes the transposed matrix of  $O$)
\begin{equation} 
J = O^{-1}DO,\quad {\rm with} \quad O^{-1}=\tilde{O}.
\end{equation}
The elements of the diagonal matrix $D$ are all positive and are chosen under the form $d_i = m \omega_i^2/2$.

The last step is an ultimate change of conjugate variables
\begin{equation}
\bm z_l = \sum^{{\cal N}-1}_{j=1} O_{lj} \bm y_j \quad \textrm{and} \quad \bm \sigma_l = \sum^{{\cal N}-1}_{j=1} O_{lj} \bm \rho_j.
\end{equation}
Expressed with these new variables, the original Hamiltonian (\ref{honr}) appears to be the sum of ${\cal N}-1$ decoupled harmonic oscillators
\begin{equation}
H_{{\rm ho}}=\sum^{{\cal N}-1}_{i=1} \left[ \frac{\bm \sigma_i^2}{2m} + \frac{1}{2} m \omega_i^2 \bm z_i^2 \right],
\end{equation}
and consequently the energy of the system is given by
\begin{equation}
\label{Eohgen}
E_{{\rm ho}}=\sum^{{\cal N}-1}_{i=1}  \omega_i (2n_i+l_i+3/2),
\end{equation}
where $n_i$ and $l_i$ are respectively the radial and orbital quantum numbers associated to the coordinate $\bm z_i$. The problem is now completely solved. Moreover, this result is valid for any excited state. Even if the expression of $\omega_i$ is in general not analytical for ${\cal N}>5$, the result (\ref{Eohgen}) is exact and could be calculated with a high accuracy. Notice that formula~(\ref{Eohgen}) extends previous results~\cite{hallho,ma}, where an equivalent mass formula is obtained in the case $k_i=0$.

\subsection{The 3-body harmonic oscillator}

The application of the general formulas to the special case ${\cal N}=3$ is very interesting: A very accurate numerical solution of the 3-body problem is now commonly obtainable and the harmonic oscillator is an obvious test of the codes. Our formulas could then be compared to numerical solutions relative to the 3-body problem. In this case the $J$-matrix is a 2 by 2 matrix so that the eigenenergies can easily be obtained. Moreover the original Hamiltonian does not favor a given particle and the eigenenergies (\ref{Eohgen}) must reflect this property; in contrast, all the formulas given in the previous section rely on the definition of the Jacobi coordinates which give a particular status to particle 3 with respect to  particles 1 and 2. It is then very instructive to check this symmetry property in the analytical expression of the energy in this particular situation.

Applying the method given previously, the result is obtained after a long but straightforward calculation. As before we introduce the reference mass $m$, the relative masses $\alpha_i=m_i/m$, the dimensionless quantities $\alpha_{ij}=\alpha_i+\alpha_j$, $\alpha=\alpha_1+\alpha_2+\alpha_3$, and the dynamical quantities $k_{ij}=k_i+k_j$ (not to be confused with $\bar k_{ij}$).
Defining the symmetric quantities
\begin{subequations}
\begin{eqnarray}
\label{defspd}
s&=&k_1\alpha_2\alpha_3 \alpha_{23}+k_2\alpha_1\alpha_3 \alpha_{13}+k_3\alpha_1\alpha_2 \alpha_{12} +\alpha(\bar k_{12}\alpha_3 \alpha_{12}+
\bar k_{13}\alpha_2 \alpha_{13}+\bar k_{23}\alpha_1 \alpha_{23}), \\
r&=&k_1k_2\alpha_3^2+k_1k_3\alpha_2^2+k_2k_3\alpha_1^2+\alpha^2(\bar k_{12}\bar k_{13}+\bar k_{13}\bar k_{23}+\bar k_{12}\bar k_{23})+ \nonumber\\
 & & \bar k_{12}(k_{12}\alpha_3^2+k_{3}\alpha_{12}^2)+\bar k_{13}(k_{13}\alpha_2^2+k_{2}\alpha_{13}^2)+\bar k_{23}(k_{23}\alpha_1^2+k_{1}\alpha_{23}^2), \\
\delta&=&\sqrt{s^2-4\alpha_1 \alpha_2 \alpha_3 \alpha r},
\end{eqnarray}
\end{subequations}
the eigenenergies of the system read
\begin{equation}
\label{Eoh3c}
E_{{\rm ho}}=\left[\frac{1}{m \alpha_1 \alpha_2 \alpha_3 \alpha} \right]^{1/2} \left[ (s+\delta)^{1/2}(2n_1+l_1+3/2)+(s-\delta)^{1/2}(2n_2+l_2+3/2)\right].
\end{equation}
They effectively present the property of not favoring one of the particles. Although this is not explicit, they do not depend on the reference mass $m$ anymore.

\subsection{Case of identical particles}

It is of interest to rewrite the solution in the case where all the particles are identical, implying that they have the same mass $m_i=m$ and the same string constants $k_i=k$, $\bar k_{ij}= \bar k$. Hamiltonian~(\ref{honr}) indeed reads in this case  
\begin{equation}\label{honr3}
	H_{{\rm ho}}=\sum^{{\cal N}}_{i=1}\frac{\bm p^2_i}{2m}+k\sum^{{\cal N}}_{i=1}(\bm r_i-\bm R)^2+\bar k\sum^{{\cal N}}_{i<j=1} (\bm r_i-\bm r_j)^2.
\end{equation}
Then, it is easy to see that the $J$ matrix is already diagonal from the very beginning so that its eigenvalues are analytically known (they all read $k+{\cal N}\bar k$) and, consequently, the eigenenergies of the system are also analytical. Explicitly they are given by ($\bm P = \bm 0$)
\begin{equation}\label{ehonr}
	E_{{\rm ho}}=\sqrt{\frac{2}{m}(k+{\cal N}\bar k)}\ Q	,
\end{equation}
where $Q$ is the total principal number
\begin{equation}\label{honriden}
	Q=\sum^{{\cal N}-1}_{i=1} (2n_i+l_i) + \frac{3}{2}({\cal N} - 1).
\end{equation}
One can check that Eq.~(\ref{Eoh3c}) reduces to Eq.~(\ref{ehonr}) with ${\cal N}=3$ for identical particles.

Wave function with good global quantum numbers and various symmetries can be built by appropriate linear combinations of states characterized by the same value of the global quantum number $Q$ \cite{silv85}. All symmetrized states have then the same energy than the nonsymmetrized states. For spatial wave functions completely symmetrical, the ground state is obtained for the values $n_i=l_i=0$ $\forall i$, so that the principal quantum number is simply
\begin{equation}\label{honrfond}
	Q_{\textrm{SGS}}=\frac{3}{2}({\cal N} - 1).
\end{equation}
For mixed symmetry or completely antisymmetrical spatial wave functions, the situation is much more involved. An estimation of the ground state energy for the completely antisymmetrical case can be computed by choosing different values for the quantum numbers and piling the states ($d$ identical values of the same $n_j$ and $l_j$ per state in order to take care of a possible spin-isospin-color degeneracy) up to the Fermi level. By considering only particle number insuring a saturated Fermi level (closed shell), one obtains 
\begin{equation}\label{honrfond2}
	Q_{\textrm{AGS}}=\frac{3}{4}({\cal N} - 1)(B_f+2),
\end{equation}
where $B_f=2 n_f+l_f$, the band number of the Fermi level, is the real positive solution of
\begin{equation}\label{Bfermi}
	{\cal N} - 1=\frac{d}{6}(B_f+1)(B_f+2)(B_f+3).
\end{equation}
Asymptotically, we have
\begin{equation}\label{honrfond3}
	\lim_{{\cal N} \gg 1} Q_{\textrm{AGS}}= \left(\frac{81}{32}\right)^{1/3}\frac{{\cal N}^{4/3}}{d^{1/3}}.
\end{equation}

\subsection{Semirelativistic case}

Let us now consider the semirelativistic generalization of~(\ref{honr3}), namely
\begin{equation}\label{hosr}
H_{{\rm ho}}=\sum^{{\cal N}}_{i=1}\sqrt{\bm p^2_i+m^2}+k\sum^{{\cal N}}_{i=1}(\bm r_i-\bm R)^2+\bar k\sum^{{\cal N}}_{i<j=1} (\bm r_i-\bm r_j)^2.
\end{equation}
Auxiliary fields, $\mu_i$, can be introduced in a similar way as it has been done in the previous section. One obtains
\begin{equation}\label{hosr2}
		H_{{\rm ho}}(\mu_j)=\sum^{{\cal N}}_{i=1}\left[\frac{\mu_i}{2}+\frac{\bm p^2_i+m^2}{2\mu_i}\right]+k\sum^{{\cal N}}_{i=1}(\bm r_i-\bm R)^2+\bar k\sum^{{\cal N}}_{i<j=1} (\bm r_i-\bm r_j)^2.
\end{equation}
Provided the total wave function is completely (anti)symmetrical, the optimal real auxiliary fields are all equals. A proof is given in the following section. Setting $\mu_i=\mu$, we obtain
\begin{equation}\label{hosr3}
		H_{{\rm ho}}(\mu)=\frac{{\cal N}}{2}\left(\mu+\frac{m^2}{\mu}\right)+\sum^{{\cal N}}_{i=1}\frac{\bm p^2_i}{2\mu}+k\sum^{{\cal N}}_{i=1}(\bm r_i-\bm R)^2+\bar k\sum^{{\cal N}}_{i<j=1} (\bm r_i-\bm r_j)^2,
\end{equation}
whose mass formula is directly obtained from~(\ref{ehonr}) provided $\bm P = \bm 0$
\begin{equation}\label{mhosr}
	M_{{\rm ho}}(\mu)=\frac{{\cal N}}{2}\left(\mu+\frac{m^2}{\mu}\right)+\sqrt{\frac{2}{\mu}(k+{\cal N}\bar k)}\ Q.
\end{equation}

The remaining auxiliary field has to be eliminated by solving $\left.\partial_\mu M_{{\rm ho}}(\mu)\right|_{\mu=\mu_0}=0$. Defining
\begin{equation}
\mu_0=\frac{2m}{\sqrt{3Y}}\, X^2\quad {\rm and}\quad Y=\frac{4m^2}{3} \left(\frac{2{\cal N}^2}{(k+{\cal N}\bar k) Q^2}\right)^{2/3},
\end{equation}
the minimization condition is given by
\begin{equation}\label{ho1}
	4X^4-8X-3Y=0.
\end{equation}
The only positive root of this last equation is given by
\begin{equation}
\label{eq:rootquarteq}
X=G_{-}(Y) = \frac{1}{2} \sqrt{v(Y)} + \frac{1}{2} \sqrt{ 4/\sqrt{v(Y)} - v(Y)},
\end{equation}
with
\begin{equation}
\label{eq:defVY}
v(Y)=\left(2 + \sqrt{4 + Y^3} \right)^{1/3} -  Y\left(2 + \sqrt{4 + Y^3}
\right)^{-1/3}.
\end{equation}
The mass formula~(\ref{mhosr}) finally becomes
\begin{eqnarray}\label{mhosr1}
M_{{\rm ho}}(\mu_0)&=&\sqrt{\frac{3}{Y}}\frac{m_t}{2 G_-(Y)^2}(4G_-(Y)+Y)=\frac{2m_t}{\sqrt{3Y}}\left[\frac{1}{G_-(Y)}+G_-(Y)^2\right].
\end{eqnarray}
Using the results of Ref.~\cite{afrela}, it can be shown that these masses are upper bounds of the eigenvalues of $H_{{\rm ho}}$. Notice the simple ultrarelativistic limit
\begin{equation}\label{mhosr2}
\lim_{m\to 0}M_{{\rm ho}}(\mu_0)=\frac{3}{2}\left[2{\cal N}(k+{\cal N}\bar k)Q^2\right]^{1/3}.
\end{equation}
It can be checked that formulas~(\ref{mhosr1}) and (\ref{mhosr2}) reduce to the two-body relations obtained in Ref.~\cite{afrela} for ${\cal N}=2$ and $k=0$. We point out that the notation $G_-(Y)$ has been chosen in order to keep the same symbols as in our previous works~\cite{af2,afrela} .

\section{Auxiliary field method}\label{afm}

\subsection{Main mass formula}\label{mainmassf}

\subsubsection{General case}

The AFM, introduced in Refs.~\cite{af,af2}, can be straightforwardly generalized to the case of ${\cal N}$-body Hamiltonians of the form~(\ref{hamgene}). The basic idea is to introduce auxiliary fields so that this Hamiltonian is formally replaced by a Hamiltonian for which an analytical solution can eventually be found. The auxiliary fields are denoted as $\mu_i$, $\nu_i$, and $\bar \nu_{ij}$, and are introduced as follows   
\begin{eqnarray}\label{hamgene2}
H(\mu_k,\nu_k,\bar\nu_{kl})&=&\sum^{{\cal N}}_{i=1}\left[\frac{\mu_i}{2}+\frac{m^2_i}{2\mu_i}+\frac{\bm p^2_i}{2\mu_i}\right]+\sum^{{\cal N}}_{i=1} \left[\nu_i\, P(r_i)+V_i(I_i(\nu_i))-\nu_i\, P(I_i(\nu_i))\right]\nonumber\\
&&+\sum^{{\cal N}}_{i<j=1} \left[\bar\nu_{ij}\, \bar P(r_{ij})+\bar V_{ij}(\bar I_{ij}(\bar \nu_{ij}))-\bar\nu_{ij}\, \bar P(\bar I_{ij}(\bar\nu_{ij}))\right]	,
\end{eqnarray}
where (notice the definition of $r_i \ne |\bm r_i|$)
\begin{equation}
r_i=|\bm r_i-\bm R|,\quad r_{ij}=|\bm r_i-\bm r_j|,	
\end{equation}
 and where
\begin{equation}\label{idef}
I_i(x)=K^{-1}_i(x),	\quad  K_i(x)=\frac{V'_i(x)}{P'(x)},\quad
\bar I_{ij}(x)=\bar K^{-1}_{ij}(x),\quad \bar K_{ij}(x)= \frac{\bar V'_{ij}(x)}{\bar P'(x)},
\end{equation}
where the prime denotes the derivative with respect to the argument.
It is sufficient to set $\mu_k=m_k$ $\forall k$ to treat a nonrelativistic kinematics.
We stress that, in order for the AFM to apply, it is assumed that the mass spectrum of~(\ref{hamgene2}) is analytically computable. In practice, this is only possible for the choice $P(x)=\bar P(x)=x^2$ under some conditions as explained in Sec.~\ref{nboh}.

The reason of such a definition is that Hamiltonians~(\ref{hamgene}) and (\ref{hamgene2}) are equivalent if the auxiliary fields are properly eliminated. It can indeed be checked that 
\begin{eqnarray}
\label{elim}
\left.\delta_{\mu_i} H(\mu_k,\nu_k,\bar\nu_{kl})\right|_{\mu_i=\hat\mu_i}=0&\Rightarrow& \hat\mu_i=\sqrt{\bm p^2_i+m^2_i}, \\
\label{elim2}
\left.\delta_{\nu_i} H(\mu_k,\nu_k,\bar\nu_{kl})\right|_{\nu_i=\hat\nu_i}=0&\Rightarrow& \hat\nu_i=K_i(r_i), \\
\label{elim3}
\left.\delta_{\bar\nu_{ij}} H(\mu_k,\nu_k,\bar\nu_{kl})\right|_{\bar\nu_{ij}=\hat{\bar\nu}_{ij}}=0&\Rightarrow& \hat{\bar\nu}_{ij}=\bar K_{ij}(r_{ij}),
\end{eqnarray}
and finally that $H(\hat\mu_k,\hat\nu_k,\hat{\bar\nu}_{kl})=H$. 

The approximation underlying the AFM is now that, as we have done in previous sections, the auxiliary fields will be seen as real variational parameters. Let us note 
\begin{equation}
H(\mu_{k,0},\nu_{k,0},\bar\nu_{kl,0}) |\varphi_0\rangle = M_0 \,|\varphi_0\rangle,  	
\end{equation}
where $\mu_{k,0}$, $\nu_{k,0}$, $\bar\nu_{kl,0}$ are the real values of the auxiliary fields such that $M_0$ is extremal. $|\varphi_0\rangle$ is an eigenstate with fixed quantum numbers and a given symmetry. Using the same mathematical techniques than in Refs.~\cite{af,afmenv,afrela}, one can show that
\begin{eqnarray}
\mu_{k,0}&=&\sqrt{\langle\varphi_0|\bm p_i^2+m_i^2|\varphi_0\rangle} \quad \textrm{with} \quad \bm P=\bm 0, \\
\langle\varphi_0|P(r_i)|\varphi_0\rangle&=&P(r_{i,0}) \quad \textrm{with} \quad r_{i,0}=I_i(\nu_{i,0}), \\
\langle\varphi_0|\bar P(r_{ij})|\varphi_0\rangle&=&\bar P(r_{ij,0}) \quad \textrm{with} \quad r_{ij,0}=I_{ij}(\bar \nu_{ij,0}).
\end{eqnarray}
It is also expected that $\langle\varphi_0|\hat \mu_i|\varphi_0\rangle \approx \mu_{i,0}$, 
$\langle\varphi_0|\hat \nu_i|\varphi_0\rangle \approx \nu_{i,0}$, and
$\langle\varphi_0|\hat{\bar\nu}_{ij}|\varphi_0\rangle \approx \bar \nu_{ij,0}$. 
Let us rewrite the Hamiltonian~(\ref{hamgene2}) under the form 
\begin{equation}
H(\mu_k,\nu_k,\bar\nu_{kl}) = T(\mu_k) + \sum^{{\cal N}}_{i=1} U_i(\nu_i,r_i) + \sum^{{\cal N}}_{i<j=1} \bar U_{ij}(\bar \nu_i,r_{ij}),
\end{equation}
where $T(\mu_k)$ stands for the kinetic part. One can also check that
\begin{align}
&U_i(\nu_{i,0},r_{i,0}) = V(r_{i,0}), \quad U'_i(\nu_{i,0},r_{i,0}) = V'(r_{i,0}), \\
&\bar U_{ij}(\bar \nu_{ij,0},r_{ij,0}) = \bar V(r_{ij,0}), \quad \bar U'_{ij}(\bar \nu_{ij,0},r_{ij,0}) = \bar V'(r_{ij,0}).
\end{align}
This means that the approximate potential $U_i$ ($\bar U_{ij}$) and the corresponding genuine potential $V_i$ ($\bar V_{ij}$) are tangent at, at least, one point.
If the following conditions $U_i(\nu_{i,0},r_i) \ge V(r_i)$ and $\bar U_{ij}(\bar \nu_{ij,0},r_{ij}) \ge \bar V(r_{ij})$ are fulfilled $\forall i$ and $\forall j$ and for all values of the radial argument, $M_0$ is an upper bound of an eigenstate of $H$ (\ref{hamgene}). If the kinematics is nonrelativistic ($T=T(\mu_k=m_k)$) and if the following conditions $U_i(\nu_{i,0},r_i) \le V(r_i)$ and $\bar U_{ij}(\bar \nu_{ij,0},r_{ij}) \le \bar V(r_{ij})$ are fulfilled $\forall i$ and $\forall j$ and for all values of the radial argument, $M_0$ is a lower bound of an eigenstate of $H$. This restriction about $T$ comes from the fact that the replacement of the genuine relativistic kinetic operator by the form $T(\mu_k)$ yields upper bounds of the eigenvalues of the genuine Hamiltonian. In the other cases, it is not possible to obtain a relevant information about the position of $M_0$ \cite{afmenv,afrela}.

At this stage, approximate numerical solutions of the ${\cal N}$-body problem can be easily computed. First, the choice $P(x)=\bar P(x)=x^2$ allows a precise determination of the eigenvalues of Hamiltonian~(\ref{hamgene2}) [see Sec.~\ref{nboh}]. Second, the extremization of the eigenvalues  with respect to the real auxiliary fields is a classical numerical problem which can be solved with a high accuracy.

\subsubsection{Identical particles}
\label{idpart}

To obtain analytical closed solutions of the eigenvalue and extremization problems associated to Hamiltonian~(\ref{hamgene2}), it is necessary to simplify the system. First of all, we will only consider systems with identical particles, that is with $m_i=m$. In this case, it is reasonable to consider identical interactions between them, namely $V_k(x)=V(x)$ and $\bar V_{kl}(x)=\bar V(x)$. This means that $K_k(x)=K(x)$, $I_k(x)=I(x)$, $\bar K_{kl}(x)=\bar K(x)$, and $\bar I_{kl}(x)=\bar I(x)$. Let us denote $\hat P_{ij}$ the permutation operator exchanging particles $i$ and $j$. We can write $\hat P_{ij}|\varphi_0\rangle =\pm |\varphi_0\rangle$ if this state $|\varphi_0\rangle$ is completely (anti)symmetrical. Then
\begin{equation}
P(r_{j,0}) = \langle\varphi_0|P(r_j)|\varphi_0\rangle = \langle\varphi_0|\hat P_{ij}P(r_i)\hat P_{ij}|\varphi_0\rangle
= \langle\varphi_0|P(r_i)|\varphi_0\rangle = P(r_{i,0}).
\end{equation}
If $P(x)$ is monotonic, which is always the case in practice, then $r_{j,0}=r_{i,0}$. Finally, if $I(x)$ is invertible, which must be the case to solve the problem, we have $\nu_{i,0}=K(r_{i,0})=K(r_{j,0})=\nu_{j,0}$. So all optimal values $\nu_{i,0}$ are the same. Using the same reasoning, we can draw the same conclusion for other auxiliary fields.

Under these conditions, we can set $\mu_i=\mu$, $\nu_i=\nu$, and $\bar \nu_{ij}=\bar\nu$ in the expression of the eigenenergies. Relations~(\ref{elim})-(\ref{elim3}) clearly show that they are related to the kinetic ($\mu_i$) and potential energies ($\nu_i$, $\bar \nu_{ij}$) in the system. We are thus led to replace the original Hamiltonian (\ref{hamgene2}) by the the following simpler Hamiltonian, which now depends on only 3 auxiliary fields
\begin{eqnarray}\label{hamgene3}
	H(\mu,\nu,\bar\nu)&=&\frac{{\cal N}}{2}\left[\mu+\frac{m^2}{\mu}\right]+{\cal N}\left[V(I(\nu))-\nu\, I(\nu)^2\right]+\frac{{\cal N}({\cal N}-1)}{2}\left[\bar V(\bar I(\bar \nu))-\bar \nu\, \bar I(\bar \nu)^2\right]\nonumber\\
	&&+\sum^{{\cal N}}_{i=1}\frac{\bm p^2_i}{2\mu}+\nu\sum^{{\cal N}}_{i=1}r_i^2+\bar \nu\sum^{{\cal N}}_{i<j=1} r_{ij}^2,
\end{eqnarray}
which, by virtue of~(\ref{mhosr}), has the following mass spectrum ($\bm P = \bm 0$)
\begin{eqnarray}\label{mass}
	M(\mu,\nu,\bar\nu)&=&\frac{{\cal N}}{2}\left[\mu+\frac{m^2}{\mu}\right]+{\cal N}\left[V(I(\nu))-\nu\, I(\nu)^2\right]+\frac{{\cal N}({\cal N}-1)}{2}\left[\bar V(\bar I(\bar \nu))-\bar \nu\, \bar I(\bar \nu)^2\right]\nonumber\\
	&&+\sqrt{\frac{2}{\mu}(\nu+{\cal N}\bar \nu)}\ Q,
\end{eqnarray}
which is characterized by a high degeneracy due to the form of $Q$. The eigenstates of $H(\mu,\nu,\bar\nu)$ are built with harmonic oscillator states. As mentioned above, states with the same energy can be coupled to good quantum numbers and a given symmetry. 

The last step needed to get the final mass formula is to find the optimal values $\mu_0$, $\nu_0$, and $\bar\nu_0$ from the extremization conditions
\begin{equation}\label{minimass}
	\left.\partial_\mu M(\mu,\nu,\bar\nu)\right|_{\mu=\mu_0}=0,\ 	\left.\partial_\nu M(\mu,\nu,\bar\nu)\right|_{\nu=\nu_0}=0,\ 	\left.\partial_{\bar\nu} M(\mu,\nu,\bar\nu)\right|_{\bar\nu=\bar \nu_0}=0.
\end{equation}
Writing explicitly Eqs.~(\ref{minimass}) and after some algebra, we arrive at the mass formula
\begin{equation}\label{massfin}
		M(\mu_0,\nu_0,\bar\nu_0)={\cal N}\mu_0+{\cal N}\, V(I(\nu_0))+\frac{{\cal N}({\cal N}-1)}{2}\bar V(\bar I(\bar \nu_0)),
\end{equation}
where the auxiliary fields are solutions of 
\begin{eqnarray}\label{mudef}
	\mu_0&=&\frac{m^2}{\mu_0}+\left[\frac{2Q^2(\nu_0+{\cal N}\bar \nu_0)}{\mu_0{\cal N}^2}\right]^{1/2}, \\
\label{nudef}
	I(\nu_0)&=&\left[\frac{Q^2}{2{\cal N}^2\mu_0(\nu_0+{\cal N}\bar \nu_0)}\right]^{1/4}, \\
\label{mubdef}
	\bar I(\bar \nu_0)&=&\left[\frac{2Q^2}{({\cal N}-1)^2\mu_0(\nu_0+{\cal N}\bar \nu_0)}\right]^{1/4}.
\end{eqnarray}
At this stage, functions $I$ and $\bar I$ are not known. They depend on the specific forms of $V$ and $\bar V$. Optimal values of $\mu_0$, $\nu_0$, and $\bar \nu_0$ are no longer known. They also depend on the state considered through the variable $Q$. Nevertheless, formula~(\ref{massfin}) makes clearly appear the mean-field nature of the AFM, because $\mu_0$ can be interpreted as the average kinetic energy of one particle, while $V(I(\nu_0))$ and $\bar V(\bar I(\bar \nu_0))$ can be respectively seen as the average potential energy of one particle in the potential $V(r)$ and of a pair in the potential $\bar V(r)$. 

\subsubsection{General remarks}\label{s4a}
A general remark about the AFM can now be done: The analytical results obtained for a given potential $V(r)$ approximated by a potential $P(r)$ can be used as a starting point for finding analytical results for an other potential $W(r)$, approximated this time by $V(r)$. This considerably enlarges the domain of applicability of this method. 

Notice that the idea of rewriting a ${\cal N}$-body Hamiltonian with pairwise interactions of the form $g(r_{ij}^2)$ as a ${\cal N}$-body harmonic oscillator has already been investigated in Ref.~\cite{hallsr2} within the framework of envelope theory. That method shares many similarities with the AFM, as shown in Ref.~\cite{afmenv}, but the results presented hereafter have, to our knowledge, never been obtained so far. An interesting result coming from the comparison between AFM and envelope theory is that it allows to gain informations about the (anti)variational nature of the mass formulas obtained. Let us indeed assume a ${\cal N}$-body Hamiltonian with pairwise interactions of the form $V_{ij}(r_{ij})=g(r_{ij}^2)$, where $g(x)$ has a definite convexity. Then, provided that the kinetic term is exactly taken into account (\emph{i.e.} in the nonrelativistic case), the energy spectrum obtained is an upper (lower) bound of the exact spectrum if $g''<0$ ($>0$). For semirelativistic Hamiltonians however, an upper bound of the kinetic term is obtained because of the introduction of auxiliary fields. Consequently, in the semirelativistic case, the AFM leads thus to upper bounds of the exact spectrum if $g''<0$. If $g''>0$ no conclusion can be drawn.

\subsection{Simplified formulations}

The four equations~(\ref{massfin})--(\ref{mubdef}) give the AFM approximation for the mass spectrum of a ${\cal N}$-body system where the constituent particles are identical. They can be even simplified by defining
\begin{equation}\label{defX0}
	X_0=\sqrt{2\mu_0(\nu_0+{\cal N}\bar \nu_0)}.
\end{equation}
We have indeed
\begin{eqnarray}\label{massfins}
		M(X_0)&=&{\cal N}\sqrt{m^2+\frac{Q}{{\cal N}}X_0}+{\cal N}\, V\left(\sqrt{\frac{Q}{{\cal N}X_0}}\right)+\frac{{\cal N}({\cal N}-1)}{2}\bar V\left(\sqrt{\frac{2Q}{({\cal N}-1)X_0}}\right),
\end{eqnarray}
where the remaining auxiliary field is a solution of
\begin{equation}\label{x0def}  
	X_0^2=2\sqrt{m^2+\frac{Q}{{\cal N}}X_0} \left[K\left(\sqrt{\frac{Q}{{\cal N}X_0}}\right)+{\cal N}\, \bar K\left(\sqrt{\frac{2Q}{({\cal N}-1)X_0}}\right)\right],
\end{equation}
and where $K$ and $\bar K$ are defined by (\ref{idef}). This condition ensures that $M(X_0)$ is extremal. The main technical difficulty is to find a closed analytical form for $M(X_0)$. Fortunately, as we shall see in the next sections, this task can be achieved in several cases of interest. It is worth mentioning that the AFM results~(\ref{massfins}) and (\ref{x0def}) are valid not only for the ground state but also for excited states. This is a serious advantage with respect to the two-body reduction presented in Sec.~\ref{twobr}. Identifying Eqs.~(\ref{massfin}) and (\ref{massfins}), it appears that
\begin{equation}
\mu_0=\sqrt{m^2+\frac{Q X_0}{{\cal N}}}.
\end{equation}
Moreover, the quantities $I(\nu_0)$ and $\bar I(\bar \nu_0)$ are directly obtained as functions of $X_0$.

\subsection{Ultrarelativistic and nonrelativistic limits}\label{subsec:ultlimgen}

For particles with null mass (ultrarelativistic limit), one obtains even simpler formulas by simply setting $m=0$ in formulas~(\ref{massfins}) and (\ref{x0def}).

In the nonrelativistic limit, the auxiliary field $\mu$ tends towards $m$. In this case also the various formulas look simpler but cautions must be taken in the limit. Explicitly, Eq.~(\ref{massfins}) reduces to
\begin{eqnarray}\label{massnrfins}
	M(X_0)&=&m_t + {\cal N}\, V\left(\sqrt{\frac{Q}{{\cal N}X_0}}\right)+\frac{Q}{2m}X_0+\frac{{\cal N}({\cal N}-1)}{2}\bar V\left(\sqrt{\frac{2Q}{({\cal N}-1)X_0}}\right),
\end{eqnarray}
where the remaining auxiliary field is a solution of
\begin{equation}\label{x0nrdef}
	X_0^2=2 \, m \left[K\left(\sqrt{\frac{Q}{{\cal N}X_0}}\right)+{\cal N}\, \bar K\left(\sqrt{\frac{2Q}{({\cal N}-1)X_0}}\right)\right].
\end{equation}

\section{Power-law potentials}\label{sec:plpot}

\subsection{General results}\label{subsec:genresplpot}

The first explicit example that will be considered below is the case of power-law potentials, \textit{i.e.} the Hamiltonian
\begin{equation}
	H_{{\rm pl}}=\sum^{{\cal N}}_{i=1}\sqrt{\bm p^2_i+m^2}+a\ {\rm sgn}(\lambda)\sum^{{\cal N}}_{i=1} r^\lambda_i+b\ {\rm sgn}(\eta)\sum^{{\cal N}}_{i<j=1} r^\eta_{ij},
\end{equation}
where ${\rm sgn}(\rho\ne 0)=|\rho|/\rho$ and where $\lambda$, $\eta\ge -1$ (in the nonrelativistic case, one can consider $\lambda$, $\eta>-2$). When only a one-body or a two-body interaction is present, parameters $a$ or $b$ must be positive. If both types of potentials are present, there are less constraints on the sign provided that a bound state could exist. Following the definitions (\ref{idef}), it is readily computed that 
\begin{equation}
	K(x)=\frac{a\, |\lambda|}{2}\, x^{\lambda-2},\quad \bar	K(x)=\frac{b\, |\eta|}{2}\, x^{\eta-2} .
\end{equation}
Then, by defining
\begin{equation}
	A_\lambda=a|\lambda|\left(\frac{{\cal N}}{Q}\right)^{\frac{2-\lambda}{2}},\quad B_\eta=b|\eta|{\cal N}\left(\frac{{\cal N}-1}{2Q}\right)^{\frac{2-\eta}{2}},
\end{equation}
Eqs.~(\ref{massfins}) and (\ref{x0def}) can be recast under the form 
\begin{eqnarray}\label{masspl}
	M(X_0)&=&{\cal N}\sqrt{m^2+\frac{Q}{{\cal N}}X_0}+Q \left(\frac{A_\lambda}{\lambda} X_0^{-\lambda/2}+\frac{B_\eta}{\eta} X_0^{-\eta/2}\right), \\
\label{xodef}
	X_0^2&=&\sqrt{m^2+\frac{Q}{{\cal N}}X_0} \left[A_\lambda\ X_0^{\frac{2-\lambda}{2}}+B_\eta\ X_0^{\frac{2-\eta}{2}}\right].
\end{eqnarray}
The sufficient condition to get a closed analytical formula for $M(X_0)$ is to solve analytically~(\ref{xodef}). This is possible if this last equation can be rewritten as a polynomial equation of the fourth degree at most. If $\lambda\neq\eta$, it can be computed that the following couples will lead to such an analytically solvable equation
\begin{eqnarray}\label{etalam}
	(\lambda,\eta)\ {\rm or}\ (\eta,\lambda)&=&(-\frac{1}{2},-1),\, (0,-1),\, (1,-1),\, (-1,-2),\, (0,-2),\, (-\frac{1}{2},-\frac{3}{2})\, .
\end{eqnarray}
Note that the last three cases are only allowed with a nonrelativistic kinematics. 

\subsection{Equal powers}\label{subsec:eqpwplpot}

\subsubsection{General case}

The problem actually becomes particularly simple when $\lambda=\eta$. In this case, the following values lead to an analytical solution  
\begin{eqnarray}\label{lamval}
	\lambda&=&-1,\, -\frac{2}{3},\, -\frac{1}{2},\, 0,\, 1,\, 2,\, -2,\, -\frac{7}{4},\, -\frac{5}{3},\, -\frac{3}{2},\, -\frac{4}{3},\, -\frac{5}{4}.
\end{eqnarray} 
The last six values are only allowed in the case of a nonrelativistic kinematics.
The solution of~(\ref{xodef}) reads
\begin{equation}\label{x00}
		X_0^{\lambda+2}=\left(A_\lambda+B_\lambda\right)^2 \left(m^2+\frac{Q}{{\cal N}}X_0\right) ,
\end{equation}
and the mass formula~(\ref{masspl}) becomes
\begin{equation}\label{M00}
	M(X_0)=\frac{{\cal N}\lambda m^2+Q (\lambda+1)X_0}{\lambda\sqrt{m^2+\frac{Q}{{\cal N}}X_0}}.
\end{equation}
Among the values~(\ref{lamval}), three cases are of obvious physical interest: $\lambda=-1$, 1, and 2. The case $\lambda=2$ corresponds to the harmonic oscillator and has been solved in Sec.~\ref{nboh}. The Coulomb problem, \textit{i.e.} $\lambda=-1$, will be specifically considered in Sec.~\ref{atom}, while we will discuss the case $\lambda=1$ hereafter. Notice that closed mass formulas for any $\lambda$ can be obtained in the nonrelativistic and ultrarelativistic limits, as we also show in the following.

\subsubsection{Ultrarelativistic and nonrelativistic limits}

In the case $m=0$, Eq.~(\ref{x00}) allows us to extract the $X_0$ value for arbitrary power
\begin{equation}
\label{plpotx0}
X_0 = \left(\frac{Q}{{\cal N}} (A_\lambda + B_\lambda)^2\right)^{\frac{1}{\lambda+1}} .
\end{equation}
Inserting this value in (\ref{M00}) provides us with the expression of the energy in the ultrarelativistic limit
\begin{equation}
\label{plpoten0}
M(X_0)=\frac{\lambda+1}{\lambda} \left[Q^{\lambda+2} {\cal N}^{\lambda} (A_\lambda + B_\lambda)^2\right]^{\frac{1}{2(\lambda+1)}}.
\end{equation}

On the contrary, assuming a large mass, $m_t \approx M(X_0)$, it is possible to extract $X_0$ from (\ref{x00})
\begin{equation}
\label{plpotxnr}
X_0=[m (A_\lambda + B_\lambda)]^{\frac{2}{\lambda+2}}.
\end{equation}
Except the Coulomb case ($\lambda=-1$) which will be treated subsequently, the term $m^2$ is dominant with respect to $X_0$. Expanding the expression of the energy (\ref{M00}) at lower order in $1/m$ leads to the nonrelativistic value of the energy
\begin{equation}
\label{plpotennr}
M(X_0)=m_t+\frac{\lambda+2}{2 \lambda} Q \left[\frac{(A_\lambda + B_\lambda)^2}{m^\lambda}\right]^{\frac{1}{\lambda+2}}.
\end{equation}

\subsubsection{Linear potential}
\label{sec:linpot}

The case $\lambda=1$ corresponds to a linearly rising confining potential. It is of great interest in hadronic physics since a linearly rising potential appears to be the best way of modeling the QCD confining interaction within potential approaches, see for example Ref.~\cite{qcd2} for more details. For the three-body problem, the one-body potential (term in $a$) corresponds to the so-called Y-junction, while the two-body potential (term in $b$) corresponds to the so-called $\Delta$ approximation. For physical problems, one has the choice of retaining one approximation or the other or both. Here for sake of generality we present the general case including both. For a given baryonic situation, one must take ${\cal N}=3$.

For $\lambda=1$, one can write
\begin{equation}
\label{abc}
A_1 + B_1 = \sqrt{\frac{{\cal N}}{Q}} c,\quad {\rm with}\quad c=a + b \sqrt{\frac{{\cal N}({\cal N}-1)}{2}}.
\end{equation}
The solution of (\ref{x00}) when $\lambda=1$ is given by 
\begin{equation}
X_0= \frac{c}{\sqrt{3}} F_-(Y),
\end{equation}
where $F_-(Y)$ is the real root of the equation $x^3-3x-2Y=0$, that is
\begin{align}\label{eq:rootcubeq}
	F_-(Y)&=\left[Y + \sqrt{Y^2- 1} \right]^{1/3} + \left[Y + \sqrt{ Y^2-1} \right]^{-1/3}\ &\rm{for}\ Y \geq 1, \nonumber \\
	  & =  2 \cos \left(\frac{1}{3}\arccos Y\right)\ &\rm{for}\ Y < 1,
\end{align}
and where
\begin{equation}
	Y=\frac{3^{3/2}{\cal N}m^2}{2 Q c}.
\end{equation}
As for $G_-(Y)$, the notation $F_-(Y)$ is chosen so that it remains consistent with our previous work~\cite{afrela}.
Introducing this value in the expression of the energy (\ref{M00}), a simple calculation allows us to obtain the energy under the form
\begin{equation}
\label{energNlin}
	M(X_0)= {\cal N} m \sqrt{\frac{F_-(Y)}{2Y}}\left[F_-(Y)+\frac{3}{F_-(Y)}\right].
\end{equation}

In the ultrarelativistic limit, where $m=0$, one obtains, from (\ref{plpoten0}), the very simple relation
\begin{equation}\label{mlin}
M(X_0)^2=4{\cal N}  c\, Q.
\end{equation}
Such a linear behavior of the square mass versus the principal quantum number is a well-known fact in hadronic physics, where light mesons and baryons are known from experimental data to exhibit Regge trajectories.

In the nonrelativistic limit, the mass formula~(\ref{plpotennr}) can be recast under the form
\begin{equation}
M(X_0) = m_t + \frac{3}{2} \left(\frac{{\cal N} Q^2 c^2}{m}\right)^{1/3}.
\end{equation}
The second term is the ``binding energy", which is a positive quantity in case of a positive linear potential.

In the special case of a linear potential, there is a peculiar relationship between the mass of the two-body problem and the mass of the ${\cal N}$-body problem. More precisely, let us call $M^{(2)}(\sigma,m,b,Q_2)$ the AFM eigenvalues of the Hamiltonian
\begin{equation}
H^{(2)} = \sigma \sqrt{\bm{p}^2+m^2}+br
\end{equation}
calculated with the natural quantum number $Q_2=2n+l+3/2$. Then, using Eq.~(\ref{energNlin}) and its counterpart for ${\cal N}=2$ in Ref.~\cite{afrela} with the value of the quantum number $Q$ defined in (\ref{honriden}), the mass $M^{(\cal N)}({\cal N},m,a,b,Q)$ of the original Hamiltonian is given by
\begin{equation}
\label{lienlinN2}
M^{(\cal N)}({\cal N},m,a,b,Q) = M^{(2)}\left({\cal N},m,a + b \sqrt{\frac{{\cal N}({\cal N}-1)}{2}},Q\right).
\end{equation}

\section{Baryonic-like systems}\label{baryon}

In this section, we want to discuss a special situation which can have an immediate application for hadronic systems, especially baryons (see Sec.~\ref{appbar}) or glueballs. Let us consider a one-body potential with a linear shape and a two-body potential of Coulomb type ($\lambda=1$ and $\eta=-1$ in subsection~{\ref{subsec:genresplpot}}). We noticed that an analytical solution does exist, but one needs to solve a general polynomial of degree 3 and the corresponding solution is rather involved. We prefer to give here the solution in the ultrarelativistic limit ($m=0$). In this special case the solution of (\ref{xodef}) is quite simple
\begin{equation}
\label{x0had}
X_0 = \frac{a}{1-b \frac{{\cal N}-1}{2Q}\sqrt{\frac{{\cal N}({\cal N}-1)}{2}}},
\end{equation}
and (\ref{masspl}) can be simplified as follows
\begin{equation}
\label{m0had}
M(X_0)=2 \sqrt{a} \sqrt{Q {\cal N} - b\left( \frac{{\cal N}({\cal N}-1)}{2}\right)^{3/2}}.
\end{equation}
Since the argument of the square root must be positive, it is readily checked that, either there exists a maximal allowed number of particles for a fixed value of $b$, either there exists a maximal allowed value of $b$ for a fixed number of particles. The relevance and the accuracy of Eq.~(\ref{m0had}) will be tested in Sec.~\ref{appbar}.

\section{Atomic systems}\label{atom}

We turn now our attention to atom-like systems, \textit{i.e.} systems described by the following Hamiltonian 
\begin{equation}\label{hat}
	H_{{\rm at}}=\sum^{{\cal N}}_{i=1}\sqrt{\bm p^2_i+m^2}-\alpha\sum^{{\cal N}}_{i=1}\frac{1}{r_i}+\bar\alpha \sum^{{\cal N}}_{i<j=1}\frac{1}{r_{ij}}.
\end{equation}
For this section, we remain completely general, the only hypothesis being that the parameters $\alpha$ and $\bar \alpha$ should be real positive numbers. The obvious application could correspond in a first approximation to ${\cal N}$ identical electrons, with charge $e$, feeling the attraction of a central static source (the nucleus) and their own repulsion. In this particular situation one must make the identification $4\pi \alpha = {\cal N} e^2$ and $4\pi \bar \alpha = e^2$. At this stage, we point out that our formalism is spin-independent and thus that this model cannot be directly applied to a real physical atom. Moreover, one should be very careful in applying the formulas already given. Two cautions are in order:
\begin{itemize}
\item In Eq.~(\ref{hat}), $r_i$ must be the distance between the particle $i$ and the center of mass of the system of ${\cal N}$ particles. In the atomic interpretation, this center of mass must coincide with the nucleus. This is not always the case but this prescription should be valid in case of a spherical atom.
\item Electrons are fermions and the total wave function (space and spin) must be completely antisymmetrical. Since the spatial wave function cannot be completely symmetrical, all possible values are not allowed for $Q$ given by Eq.~(\ref{honriden}). For closed shell atoms, $Q_{\textrm{AGS}}$ (\ref{honrfond2}) should be used to estimate the ground state energy only.
\end{itemize}

Nevertheless, solving Hamiltonian~(\ref{hat}) has an intrinsic interest since, to our knowledge, no corresponding analytical mass formula is known so far. A general solution can moreover be found with the relativistic kinematics, starting from~(\ref{x00}) and (\ref{M00}). Applying them to the case $\lambda = -1$, $a = \alpha$, $b = - \bar \alpha$ and defining
\begin{equation}
\label{defDat}
D=\frac{1}{Q} \left[\alpha {\cal N} - \frac{\bar \alpha}{{\cal N}} \left(\frac{{\cal N}({\cal N}-1)}{2}\right)^{3/2}\right],
\end{equation}
the solution of equation~(\ref{x00}) is given by
\begin{equation}
	X_0=\frac{{\cal N} m^2 D^2}{Q(1-D^2)}.
\end{equation}
The mass formula~(\ref{M00}) then reads
\begin{equation}\label{matsr}
M(X_0) = m_t \sqrt{1 - D^2}.
\end{equation}

In the ultrarelativistic limit, this expression gives the value $M = 0$. This property is well known in the two-particle system and prolongates to the ${\cal N}$-body problem, at least at this level of approximation. Nevertheless, it is quite obvious that a system of massless particles cannot be characterized by a (negative) binding energy since the resulting mass would be negative. Moreover, with a Coulomb-type potential, the only energy scale of the problem is the particle mass. So, if this mass vanishes, no bound state can exist. In the nonrelativistic regime, one must assume that the $D$ quantity is small in Eq.~(\ref{matsr}). In this case
\begin{equation}
	M^{{\rm nr}}= m_t - \frac{1}{2} m_t D^2.
\end{equation}

It is clear from formula~(\ref{matsr}) that the value of the parameters cannot be arbitrary. One must satisfy the condition $D < 1$. Explicitly this means
\begin{equation}
\label{condD}
\alpha {\cal N}^2 - \bar \alpha \left(\frac{{\cal N}({\cal N}-1)}{2}\right)^{3/2} < Q {\cal N}.
\end{equation}
Let us emphasize that this condition is independent of the mass of the particle \cite{lieb88}. Let us notice that, for $4\pi \alpha = {\cal N} e^2$ and $4\pi \bar \alpha = e^2$, a maximal value of ${\cal N}$ exists if $Q$ increases less rapidly that ${\cal N}^2$.

\section{Pairwise interactions}\label{pair}

In this section, we focus on the case where the Hamiltonian of the system only contains pairwise interactions. This is relevant in many physical ${\cal N}$-body problems. Of course, all the formulae given in the previous sections where we considered a mixing of one- and two- body interactions remain valid; it is sufficient to set $V(r)=0$ and $K(r)=0$ in them. But we have additional interesting properties in this case.

\subsection{Gaussian potential}
\label{subsec:gauspot}

We have considered, up to now, potentials which admit an infinite number of bound states. There are however many cases in which the number of bound states is finite. One family of such potentials are of the form $-\alpha\, r^{\lambda}\, {\rm e}^{-\beta\, r^{\eta}}$. Solutions of the Schr\"odinger equation with those potentials have been found by using the AFM in Ref.~\cite{af3} for $\eta=1$. Moreover, recent results have been obtained for semirelativistic $N$-boson systems bound by attractive pair potentials and in particular for pair exponential potentials \cite{hall5}. However, Gaussian potentials are sometimes used in ${\cal N}$-body computations, typically in nuclear physics (see for example Ref.~\cite{nucl}) or Bose-Einstein condensates~\cite{bec}. That is why we now propose to study the nonrelativistic Hamiltonian
\begin{equation}\label{hgau}
	H_{{\rm ga}}=\sum^{{\cal N}}_{i=1}\frac{\bm p^2_i}{2m}-\alpha\sum^{{\cal N}}_{i<j=1}{\rm e}^{-\beta^2\, r^2_{ij}}.
\end{equation}
Let us perform the change of variables $\bm x_i=\beta\bm r_i$, $\bm q_i=\bm p_i/\beta$. Then we have
\begin{equation}
\label{htgau}
	\tilde H_{{\rm ga}}=\frac{m}{\beta^2}H_{{\rm ga}}=\sum^{{\cal N}}_{i=1}\frac{\bm q^2_i}{2}-g\sum^{{\cal N}}_{i<j=1}{\rm e}^{-x^2_{ij}},\quad {\rm with}\quad g=\frac{m\alpha}{\beta^2}.
\end{equation}
Only the potential depth $g$ is a relevant parameter of the system. 

The AFM can now be applied to solve $\tilde H_{{\rm ga}}$. We have 
\begin{equation}
	I(\nu)=\sqrt{-\ln\left(\frac{\nu}{g}\right)},
\end{equation}
and consequently the energy spectrum reads, after simplification of Eq.~(\ref{massfin}) with $\mu_0=m$ and $E=M-m_t$,
\begin{equation}
	\tilde E_{{\rm ga}}(\nu)=\sqrt{2{\cal N} \nu\, Q^2}+\frac{{\cal N}({\cal N}-1)}{2}\nu\left[\ln\left(\frac{\nu}{g}\right)-1\right].
\end{equation}
The equation of minimization of the energy with respect to $\nu$  is given by 
\begin{equation}\label{nming}
	\nu_0^{1/2}\, \ln\left(\frac{\nu_0}{g}\right)=-\sqrt{\frac{2Q^2}{{\cal N}({\cal N}-1)^2}},
\end{equation}
and its solution reads
\begin{equation}
	\nu_0=g\, {\rm e}^{2W_0(Y)} \quad {\rm with}\quad Y=-\frac{Q}{\sqrt{2g{\cal N}({\cal N}-1)^2}}.
\end{equation}
$W_0(x)$ is the principal branch of the Lambert function, defined through $W_0(x)\, {\rm e}^{W_0(x)}=x$ (see Ref.~\cite{af3} for more details). After simplification, the energy spectrum is given by
\begin{equation}
	\tilde E_{{\rm ga}}(Y)=-\frac{Q^2}{({\cal N}-1)}\frac{1+2W_0(Y)}{4W_0(Y)^2}.	
\end{equation}
In physical units, using the rescaling~(\ref{htgau}), the energy spectrum of Hamiltonian (\ref{hgau}) finally reads
\begin{equation}\label{egau}
	E_{{\rm ga}}(Y)=\frac{\beta^2}{m}\tilde E_{{\rm ga}}(Y) \quad {\rm with}\quad Y=-\frac{\beta Q}{({\cal N}-1)\sqrt{2{\cal N}m\alpha}}.
\end{equation}
Denoting explicitly by $E^{(2)}(m,\alpha,\beta)$ the previous formula applied to the two-body problem (${\cal N}=2$), it is quite easy to express the energy $E^{(\cal N)}(m,\alpha,\beta)$ of the ${\cal N}$-body problem given by Eq.~(\ref{egau}) by the duality equation
\begin{equation}
\label{dualgaus}
E^{(\cal N)}(m,\alpha,\beta) = E^{(2)}\left(m, \alpha \frac{{\cal N}({\cal N}-1)}{2}, \frac{\beta}{\sqrt{{\cal N}-1}}\right).
\end{equation}
An alternative form for the energy of the system is
\begin{equation}
\label{egausalt}
E^{(\cal N)}(m,\alpha,\beta)= -\frac{{\cal N}({\cal N}-1)}{2} \alpha Y^2 \frac{ 1+2W_0(Y)}{W_0(Y)^2}.
\end{equation}
This last formula is an upper bound of the exact energy spectrum since $g(x)=-\alpha\, {\rm e}^{-\beta^2 x}$ is such that $g''<0$ (see Sec.~\ref{s4a}). A similar upper bound on the ground state energy has already been obtained in Ref.~\cite{kinetic} by using a trial wave function of the form ${\rm e}^{-\kappa\sum_{i<j}r_{ij}^2}$. It is interesting to say that, although that trial wave function is similar to the harmonic oscillator wave function we obtain in the AFM, both results are inequivalent. The main reason is that, in such a variational computation, the scale parameter $\kappa$ is fitted so that the ground state energy is minimal. In the AFM, the scale parameter of the wave function is typically proportional to $\sqrt{m\nu_0}$, where $\nu_0$ is computed from the extremum condition~(\ref{nming}) concerning the whole energy spectrum. Thus, even in the ground state case, both results do not coincide.  

It is interesting to check that Eq.~(\ref{egau}) has the correct limit $\beta\to 0$. In the case of a small $\beta$, the potential term is approximately equal to $\sum^{{\cal N}}_{i<j=1}(-\alpha+\alpha\beta^2\, r^2_{ij})$, \textit{i.e.} a harmonic oscillator. As expected, in agreement with formula~(\ref{ehonr}), we obtain
\begin{equation}
E_{{\rm ga}}(\nu_0)= -\frac{{\cal N}({\cal N}-1)}{2}\alpha+\sqrt{\frac{2\alpha\beta^2 {\cal N}}{m}}\, Q
+{\cal O}(\beta^2)
\end{equation}
by performing an expansion of formula~(\ref{egau}) in powers of $\beta$.

It is well-known that exponential-like potentials only admit a finite number of bound states. In particular, the energy formula~(\ref{egau}) vanishes for some value of $g$. These values are called critical coupling constants since they correspond to potential strength for which a given bound state appears (or disappears). By solving the equation $1+2W_0(Y_c)=0$, one finds $Y_c=-1/(2\sqrt {\rm e})$ and consequently the critical coupling constants read
\begin{equation}
	g_{{\cal N}}=\frac{2{\rm e}Q^2}{{\cal N}({\cal N}-1)^2}.
\end{equation}
For symmetrical spatial wave function, the ground state exists only if $g>\left.g_{{\cal N}}\right|_{n_i=l_i=0}=9{\rm e}/2{\cal N}$; we obtain in this case the relation
\begin{equation}
	\left.\frac{g_{{\cal N}+1}}{g_{{\cal N}}}\right|_{n_i=l_i=0}=\frac{{\cal N}}{{\cal N}+1},
\end{equation}
that has previously been obtained and numerically checked for several exponential-type potentials~\cite{gauss}. Similarly
\begin{equation}
	\left.g_{\cal N}\right|_{n_i=l_i=0}= \frac{2}{{\cal N}} 	\left.g_2\right|_{n_i=l_i=0},
\end{equation}
indicating that in order to link a ${\cal N}$-body system one needs a coupling ${\cal N}/2$ times smaller than the coupling for a two-body problem \cite{gauss}.

\subsection{Two-body reduction revisited}

In the case of two identical particles, $r_1=r_2=r_{12}/2$. So, the total interaction is just a function of $r_{12}$ and there is no need to make a distinction between $V$ and $\bar V$ potentials. For instance, in the case of two identical power-law potentials, the total interaction $V_t$ is simply given by
\begin{equation}\label{W}
	V_t=\textrm{sgn}(\lambda) \left(2^{1-\lambda} a + b \right) r_{12}^\lambda.
\end{equation}

The mass formula coming from the application of the AFM to the ${\cal N}$-body Hamiltonian~(\ref{hamgene}) is defined through the equations~(\ref{massfin})-(\ref{mubdef}). The same procedure can actually be applied to find the spectrum of the two-body-like Hamiltonian
\begin{equation}\label{hdual}
	H^*=\sigma\sqrt{\bm p^2+m^2}+g\, \bar V(r),
\end{equation}
where $\sigma$ and $g$ are positive parameters (the $^*$ is added to denote the two-body quantities). One finds
\begin{equation}\label{massfin2}
	M^*(\mu_0^*,\bar\nu_0^*)=\sigma\mu_0^*+g\, \bar V(\bar I(\bar \nu_0^*)),
\end{equation}
where the auxiliary fields are solutions of 
\begin{eqnarray}\label{mudef2}
	\mu_0^*&=&\frac{m^2}{\mu_0^*}+\left[\frac{2g\, Q^{*2}\bar \nu_0^*}{\sigma\mu_0^*}\right]^{1/2}, \\
\label{mubdef2}
	\bar I(\bar \nu_0^*)&=&\left[\frac{\sigma\, Q^{*2}}{2g\, \mu_0^*\bar \nu_0^*}\right]^{1/4}.
\end{eqnarray}

Equations~(\ref{massfin2})-(\ref{mubdef2}) are actually equivalent to (\ref{massfin})-(\ref{mubdef}), provided that only pairwise interactions are considered and that the following substitutions are made
\begin{equation}
	\mu_0^*=\mu_0,\quad \bar\nu_0^*=\bar\nu_0,\quad 
	\frac{g}{\sigma}=\frac{{\cal N}-1}{2},\quad Q^{*2}=\frac{2Q^2}{{\cal N}({\cal N}-1)}.
\end{equation}
Then, if $M^{(2)}(\sigma,g,Q^*)$ is the the mass spectrum of the hamiltonian~(\ref{hdual}), it follows that the mass spectrum of the Hamiltonian
\begin{equation}\label{hamgenep}
	H=\sum^{{\cal N}}_{i=1}\sqrt{\bm p^2_i+m^2}+\sum^{{\cal N}}_{i<j=1}\bar V(r_{ij})
\end{equation}
reads
\begin{equation}\label{mdual2}
	M^{({\cal N})}(Q)=\frac{{\cal N}}{\sigma}\, M^{(2)}\left(\sigma,\frac{{\cal N}-1}{2}\sigma,Q\sqrt{\frac{2}{{\cal N}({\cal N}-1)}}\right),
\end{equation}
or equivalently 
\begin{equation}\label{mdual}
		M^{({\cal N})}(Q)=\frac{{\cal N}({\cal N}-1)}{2g}\, M^{(2)}\left(\frac{2g}{{\cal N}-1},g,Q\sqrt{\frac{2}{{\cal N}({\cal N}-1)}}\right).
\end{equation}
In these last two expressions, $Q$ is given by~(\ref{honriden}).

\subsection{Funnel potential}

The funnel potential $ar-b/r$ is one of the most widely used effective QCD potential in hadronic physics. It has actually been confirmed by lattice QCD as being an excellent parameterization of the potential energy between a static quark-antiquark pair~\cite{qcd2}. But, this potential is also believed to be applicable in light hadron physics, and it is of interest to find an analytical mass formula for the Hamiltonian
\begin{equation}\label{hamgenef}
	H_{{\rm f}}=\sum^{{\cal N}}_{i=1}\sqrt{\bm p^2_i}+\sum^{{\cal N}}_{i<j=1}\left[a\, r_{ij}-\frac{b}{r_{ij}}\right].
\end{equation}
It has been shown in Ref.~\cite{afrela} that the mass spectrum of 
\begin{equation}\label{h2fun}
H^{(2)}_{{\rm f}}=\sigma\sqrt{\bm p^2+m^2}+ar-b/r	
\end{equation}
is given by 
\begin{equation}\label{m2fun}
M^{(2)}_{{\rm f}}=2\sqrt{a(\sigma Q^*-b)}	
\end{equation}
within the AFM in the case $m=0$.
Applying the duality relation~(\ref{mdual}), one is led to the mass formula
\begin{equation}
	M_{{\rm f}}^2=a\sqrt{8{\cal N}({\cal N}-1)}{\cal N}\, Q-ab{\cal N}^2({\cal N}-1)^2.
\end{equation}
It is worth mentioning that this last formula can be computed without using a duality relation, but rather by applying directly Eqs.~(\ref{massfins}) and (\ref{x0def}). This constitutes a good cross-check of our formulas.

For completeness, we mention that, in Ref.~\cite{hallsr1}, various upper and lower bounds for the mass spectrum of Hamiltonian~(\ref{h2fun}) are presented. In our notations, they are mostly of the form
\begin{equation}\label{hallb}
	\bar M_{{\rm f}}=\min_{r>0}\left[\sigma\sqrt{m^2+\frac{P^2}{r^2}}+a\,  k_1\, r-\frac{b\, k_{-1}}{r}\right],
\end{equation}
where $P$, $k_1$, and $k_{-1}$ are real numbers depending on the approximation scheme which is used (the quantum numbers can be contained in $P$). In the ultrarelativistic limit, the minimization can be performed easily and one gets
	\begin{equation}
	\bar M_{{\rm f}}=2\sqrt{a\, k_1(\sigma\, P-b\, k_{-1})},
\end{equation}
which is formally identical to our formula~(\ref{m2fun}). This is another consistency check of the AFM since formulas~(\ref{hallb}) have been obtained independently by using other methods.   

\section{Application to baryons}\label{appbar}

In order to test the relevance of our method, we apply it for a particular relativistic three-body system: the light baryon composed of three massless quarks. In the framework of constituent models, the Hamiltonian for such a system is given by \cite{qcd1a,qcd1b}
\begin{equation}\label{hbaryon}
H^\textrm{B}=\sum^3_{i=1}\sqrt{\bm p^2_i}+\lambda \sum^3_{i=1} |\bm r_i-\bm R|
-\frac{2}{3}\alpha_S\sum^3_{i<j=1}\frac{1}{|\bm r_i-\bm r_j|}.
\end{equation}
The dominant interaction is a confinement by three strings with density $\lambda$ meeting at the center of mass. The short-range part is given by pairwise interactions of Coulomb nature with a strong coupling constant $\alpha_S$. In order to allow a more realistic calculation of the correct baryon masses, this Hamiltonian must be completed with terms that can be computed in perturbation \cite{qcd1a}. As these contributions represent no interest for the AFM, they are not considered in this study. 

Let us note $H^\textrm{AFB}$ the counterpart of $H^\textrm{B}$ once the auxiliary fields are introduced. The eigenstates of $H^\textrm{AFB}$ are built with harmonic oscillator states and the eigenvalues are given by the application of Eq.~(\ref{m0had}) with ${\cal N}=3$, $a=\lambda$ and $b=2\alpha_S/3$, 
\begin{equation}
\label{m0bar}
M^\textrm{AFB}_0 = \sqrt{12\lambda \left(B+3-\frac{2 \alpha_S}{\sqrt{3}}\right)},
\end{equation}
where $B$ is the band number defined as $B=2(n_1+n_2)+(l_1+l_2)$; see Eq.~(\ref{honriden}). One can notice the strong degeneracy due to the particular form of $B$. This formula generalizes a result obtained for a small value of $\alpha_S$ in Ref.~\cite{qcd1a}. Following the considerations of Sec.~\ref{mainmassf}, These masses are upper bounds of the exact masses.

In order to test the relevance of this formula, it is necessary to compute accurately eigenvalues of $H^\textrm{B}$. It is possible, for instance, to use a variational method relying on the expansion of trial states with a harmonic oscillator basis \cite{hobasis}. We can write
\begin{equation}
\label{hotrial}
|\psi\rangle=\sum_{B=0}^{B_\textrm{max}}\sum_{q(B)} |\phi(B,q(B))\rangle,
\end{equation}
where $B$ characterizes the band number of the basis state and where $q$ summarizes all the quantum numbers of the state (which can depend on $B$). This procedure is specially interesting since the eigenstates of $H^\textrm{B}$ are expanded in term of eigenstates of $H^\textrm{AFB}$ (up to a length scale factor). In practice, a relative accuracy better than $10^{-4}$ is reached with $B_\textrm{max}=20$. Such results are denoted ``exact" in the following.

Before comparing exact masses with formula~(\ref{m0bar}), it is significant to describe the baryon wave functions. Quarks are fermions of spin $1/2$ with isospin and color degrees of freedom. The global color function is unique and completely antisymmetrical. For total spin $S=3/2$ (isospin $T=3/2$) the spin (isospin) wave function is completely symmetrical. The corresponding wave function are of mixed symmetry for $S=1/2$ or $T=1/2$. Using degenerate eigenstates of $H^\textrm{AFB}$, it is always possible to build three-quarks states completely antisymmetrical \cite{silv85,hobasis} which are characterized by the same mass. For instance, when $S=T=3/2$ or $S=T=1/2$, the baryon states possess a spatial wave function completely symmetrical. In the real world, the degeneracies are removed by spin- or isospin-dependent operators which can be computed as perturbations \cite{qcd1a}. When an eigenstate of $H^\textrm{B}$ is computed for given values of $S$, $T$, and the total angular momentum $L$, we determine the band number of its dominant components. It can then be compared with the eigenstate of $H^\textrm{AFB}$ with the same spin-isospin quantum numbers and the same band number. 

In Table~\ref{tab:comp}, exact masses are compared with the predictions of formula~(\ref{m0bar}) for usual values of the parameters $\lambda$ and $\alpha_S$. The relative error can be large but the quality of the approximation is quite reasonable for a so simple formula, and masses are all upper bounds. Due to the strong degeneracy of the harmonic oscillator, all eigenstates of $H^\textrm{AFB}$ with the same band number have the same energy. This approximation is better for high values of $L$. Let us note that the relative error can be reduced by a factor around 2 when $\alpha_S\to 0$. When $\alpha_S= 0$, the relative error becomes independent of the energy scale factor $\sqrt{\lambda}$.

\begin{table}[ht]
\caption{\label{tab:comp} Exact eigenmasses of the Hamiltonian~(\ref{hbaryon}) as a function of the band number $B$ and the total angular momentum $L$ for $\lambda=0.2$~GeV$^2$ and $\alpha_S=0.4$. The number in brackets is the probability (\%) of the component with the band number $B$ in the harmonic oscillator expansion~(\ref{hotrial}). These results are compared with the masses given by formulas~(\ref{m0bar}), (\ref{m1bar}), and (\ref{m2bar}). The number in parenthesis is the relative error (\%) with respect to the exact number. All masses are given in GeV.}
\begin{ruledtabular}
\begin{tabular}{cccccc}
$B$   & $L$ & Exact & $M^\textrm{AFB}_0$~(\ref{m0bar}) & $M^\textrm{AFB}_1$~(\ref{m1bar}) & $M^\textrm{AFB}_2$~(\ref{m2bar}) \\
\hline
0 & 0 & 2.128 [92.9] & 2.468 (16.0) & 2.168 (\phantom{0}1.9) & 2.168 (1.9) \\
1 & 1 & 2.606 [95.9] & 2.914 (11.8) & 2.596 (\phantom{0}0.4) & 2.596 (0.4) \\
2 & 0 & 2.739 [89.4] & 3.300 (20.5) & 2.962 (\phantom{0}8.1) & 2.811 (2.6) \\
  & 2 & 2.959 [96.0] & 3.300 (11.5) & 2.962 (\phantom{0}0.1) & 2.962 (0.1) \\
3 & 1 & 3.125 [91.7] & 3.646 (16.7) & 3.288 (\phantom{0}5.2) & 3.152 (0.9) \\
  & 3 & 3.299 [96.7] & 3.646 (10.5) & 3.288 (\phantom{0}0.3) & 3.288 (0.3) \\
4 & 0 & 3.260 [80.8] & 3.961 (21.5) & 3.585           (10.0) & 3.332 (2.2) \\
  & 2 & 3.422 [92.3] & 3.961 (15.8) & 3.585 (\phantom{0}4.7) & 3.460 (1.1) \\
  & 4 & 3.581 [96.8] & 3.961 (10.6) & 3.585 (\phantom{0}0.1) & 3.585 (0.1) \\
5 & 1 & 3.584 [86.3] & 4.253 (18.7) & 3.858 (\phantom{0}7.7) & 3.625 (1.1) \\
  & 3 & 3.716 [93.6] & 4.253 (14.5) & 3.858 (\phantom{0}3.8) & 3.743 (0.7) \\
  & 5 & 3.861 [97.0] & 4.253 (10.2) & 3.858 (\phantom{0}0.1) & 3.858 (0.1) \\
6 & 0 & 3.721 [74.4] & 4.527 (21.7) & 4.114           (10.6) & 3.782 (1.6) \\
  & 2 & 3.838 [86.4] & 4.527 (17.9) & 4.114 (\phantom{0}7.2) & 3.895 (1.5) \\
  & 4 & 3.966 [93.6] & 4.527 (14.1) & 4.114 (\phantom{0}3.7) & 4.006 (1.0) \\ 
  & 6 & 4.103 [96.9] & 4.527 (10.3) & 4.114 (\phantom{0}0.3) & 4.114 (0.3) \\ 
\end{tabular}
\end{ruledtabular}
\end{table}

In our previous work about the AFM \cite{af,af2,af3,afmenv,afrela}, we have shown that it is possible to improve two-body mass formulas by changing the structure of the global quantum numbers $Q$. By fitting another form on exact eigenvalues, a very high accuracy can sometimes be reached. In this work, we will proceed differently and will try to use analytical results to find the best shape for the mass formula.

An upper bound of the ground state of $H^\textrm{B}$ can be computed by using the expansion~(\ref{hotrial}) with just one state for which $B_\textrm{max}=0$ (with harmonic oscillators reduced to Gaussian states). In this special case, a very simple form is obtained
\begin{equation}
\label{mbargs0}
M^\textrm{B}(B_\textrm{max}=0) = \sqrt{\frac{32}{\pi}\lambda\left( 3-\sqrt{3} \alpha_S\right)}.
\end{equation}
This result is a better upper bound of the exact ground state than the formula~(\ref{m0bar}) with $B=0$. So we can try to improve this last formula simply by the changes: $12 \to 32/\pi$ and $2/\sqrt{3} \to \sqrt{3}$. This gives 
\begin{equation}
\label{m1bar}
M^\textrm{AFB}_1 = \sqrt{\frac{32}{\pi}\lambda\left( B+3-\sqrt{3} \alpha_S\right)}.
\end{equation}
One can see in Table~\ref{tab:comp} that the masses are greatly improved, but the variational character of the formula~(\ref{m1bar}) cannot longer be guaranteed: some masses are now below the exact ones. Moreover, the problem of the degeneracy remains. To cure this situation, we will look at a similar two-body system. 

The masses of mesons composed of two massless quarks can be computed with the two-body equivalent of the Hamiltonian~(\ref{hbaryon}). In this case, for large values of the quantum numbers, it has been shown by the WKB method that the mass depend directly of the combination $\pi n/2 + l$ \cite{brau00}. These numbers are in agreement with a result obtained numerically in the Secs.~9.1 and 9.3 of Ref.~\cite{afrela}, even for small values of $n$ and $l$. For small quark masses, the long-range part of the potential dominates the dynamics of the baryon and we can assume that Eq.~(\ref{lienlinN2}) is valid for the Hamiltonian~(\ref{hbaryon}), that is to say that the $Q$-dependence is the same for two- and ${\cal N}$-body systems. So, we can try to improve formula~(\ref{m1bar}) by simply setting
\begin{equation}
\label{m2bar}
M^\textrm{AFB}_2 = \sqrt{\frac{32}{\pi}\lambda\left( B'+3-\sqrt{3} \alpha_S\right)}
\quad \textrm{with} \quad B'=\frac{\pi}{2}(n_1+n_2)+(l_1+l_2),
\end{equation}
which removes the strong degeneracy of the harmonic oscillator behavior. One can see in Table~\ref{tab:comp}, that the relative error is now around 1\%. Despite its simplicity and its nonvariational character, Eq.~(\ref{m2bar}) is then a very good mass formula for the eigenstates of the Hamiltonian~(\ref{hbaryon}). It is not sure that the procedure used here to improve the mass formula for baryons could work so well for other Hamiltonians. But, this shows that an improvement is possible, at least in some particular cases. 

\section{Conclusions}\label{conclu}

The auxiliary field method, which is strongly connected with the envelope theory \cite{afmenv}, is a powerful tool to compute approximate closed analytical solutions of the two-body Schr\"odinger equation with various types of interactions \cite{af,af2,af3,hybri}. The procedure starts with the replacement of an arbitrary potential $V(r)$ by another one $\tilde V(r) = \nu P(r)+g(\nu)$, $P(r)$ being a potential for which analytical eigenenergies can be found, $\nu$ the auxiliary field and $g(\nu)$ a well-defined function of this extra field. With a proper elimination of $\nu$, the original potential is recovered. The basic idea underlying this method is to consider the auxiliary field as a real number which is eventually eliminated by a condition rendering extremal the eigenenergies for the potential $\tilde V(r)$. This technique has been recently extended to semirelativistic two-body Hamiltonians \cite{afrela}.  

In this paper, we apply the auxiliary field method to ${\cal N}$-body problems with nonrelativistic and relativistic kinematics. As this method requires the knowledge of a solvable problem as a starting point, the system of ${\cal N}$-harmonic oscillators is first studied and solved. Some general results are given for ${\cal N}$-body systems with one-body and two-body interactions. If numerical approximate solutions can be easily computed, it is not possible to obtain closed analytical formulas for such complicated problems. So, we focus our attention to the case of ${\cal N}$ identical particles for which it is shown that the number of unknown auxiliary fields can be reduced to three or even one. Several problems are then studied within this special framework. The case of power-law one-body interactions plus power-law two-body interactions is studied and mass formulas are given for atomic-like and baryonic-like systems. Not only the ground state is computed but also all excitations. Up to our knowledge, these results for ${\cal N}$-body problems are obtained for the first time. The case of only pairwise interactions deserves a particular attention. In this case, dual formulas connecting the ${\cal N}$-body problem to the 2-body problem with other parameters are obtained. A mass formula for the funnel potential is computed, due to its significance in hadronic physics. As an application, a Hamiltonian describing a baryon composed of three massless quarks is studied. Accurate numerical eigenvalues are compared with the predictions of our method. Results obtained are quite good for the very simple mass formula computed. The formula has even been improved by using supplementary analytical results coming from variational and WKB calculations. The auxiliary field method yields also analytical approximation of the eigenstates \cite{sema10}. They are product of harmonic oscillator states if the potentials $P(x)$ and $\bar P(x)$ are quadratic ones. Moreover, in the case of identical particles, states with a given symmetry can be built by a well defined procedure \cite{silv85}.

Other types of problems could be treated by our method, for instance a chain-like configuration (phonons, gluons in a glueball, etc.). For example, the relevance of ``polymer chains" of quarks and gluons in a quark-gluon plasma has been already discussed in the literature \cite{liao06}. 

\section*{Acknowledgments}
C. Semay and F. Buisseret would thank the F.R.S.-FNRS for financial support. F. Brau acknowledges financial support from a return grant delivered by the Federal Scientific Politics.

\end{document}